\documentclass[floatfix,twocolumn,showpacs,aps,amsmath,amssymb]{revtex4}
\usepackage{psfrag}
\usepackage{graphicx}
\usepackage[labelsep=space,justification=raggedright,font=normalsize]{caption}

\usepackage{amsmath,amssymb}
\usepackage{xspace}
\newcommand{\xmax}{\ensuremath{X_\mathrm{max}}}
\newcommand{\xmaxv}{\ensuremath{X_\mathrm{max}^\mathrm{v}}}
\newcommand{\gsm}{g/cm${}^2$}
\newcommand{\dgv}{$DG^\mathrm{v}$}

\newcommand{\sigrat}{$S_\mu/S_\mathrm{EM}$}
\newcommand{\sem}{\ensuremath{S_\mathrm{EM}}}

\newcommand{\denrat}{$D_\mu/D_\mathrm{e}$}

\newcommand{\logen}{\log10(E)[\mathrm{eV}]=}

\begin{document}

\title{Precise determination of muon and electromagnetic shower
  contents from shower universality property}

\author{A.~Yushkov\footnote{yushkov@na.infn.it}}
\affiliation{INFN Sezione di Napoli, I-80126, via Cintia, Napoli, Italia}

\author{M.~Ambrosio}
\affiliation{INFN Sezione di Napoli, I-80126, via Cintia, Napoli, Italia}

\author{C.~Aramo}
\affiliation{INFN Sezione di Napoli, I-80126, via Cintia, Napoli, Italia}

\author{F.~Guarino}
\affiliation{INFN Sezione di Napoli, I-80126, via Cintia, Napoli, Italia}
\affiliation{Universit\`{a} di Napoli ``Federico~II'', I-80126, via Cintia, Napoli, Italia}

\author{D.~D'Urso}
\affiliation{INFN Sezione di Napoli, I-80126, via Cintia, Napoli, Italia}

\author{L.~Valore}
\affiliation{INFN Sezione di Napoli, I-80126, via Cintia, Napoli, Italia}

\date{\today}

\begin{abstract}
We consider two new aspects of Extensive Air Shower (EAS) development
universality allowing to make accurate estimation of muon and
electromagnetic (EM) shower contents in two independent ways. In the
first case, to get muon (or EM) signal in water Cherenkov tanks or in
scintillator detectors, it is enough to know the vertical depth of
shower maximum \xmaxv\ and the total signal in the ground detector. In the
second case, the EM signal can be calculated from the primary particle
energy and the zenith angle. In both cases the parametrizations of
muon and EM signals are almost independent on primary particle nature,
energy and zenith angle. Implications of the considered properties
for mass composition and hadronic interaction studies are briefly
discussed.

The present study is performed on 28000 of proton, oxygen and iron
showers, generated with CORSIKA~6.735 for $E^{-1}$ spectrum in the
energy range $\logen18.5-20$ and uniformly distributed in
$\cos^2{\theta}$ in zenith angle interval $\theta=0^\circ-65^\circ$
for QGSJET~II/Fluka interaction models.
\end{abstract}


\iffalse
* 96.50.sd Extensive air showers
* 96.50.sb Composition, energy spectra and interactions
* 29.40.Ka Cherenkov detectors
* 29.40.Mc Scintillation detectors
* 95.55.Vj Neutrino, muon, pion, and other elementary particle
* detectors; cosmic ray detectors
* 95.85.Ry Neutrino, muon, pion, and other elementary particles; cosmic rays
\fi

\maketitle

\begin{figure*}
\includegraphics[width=0.49\textwidth]{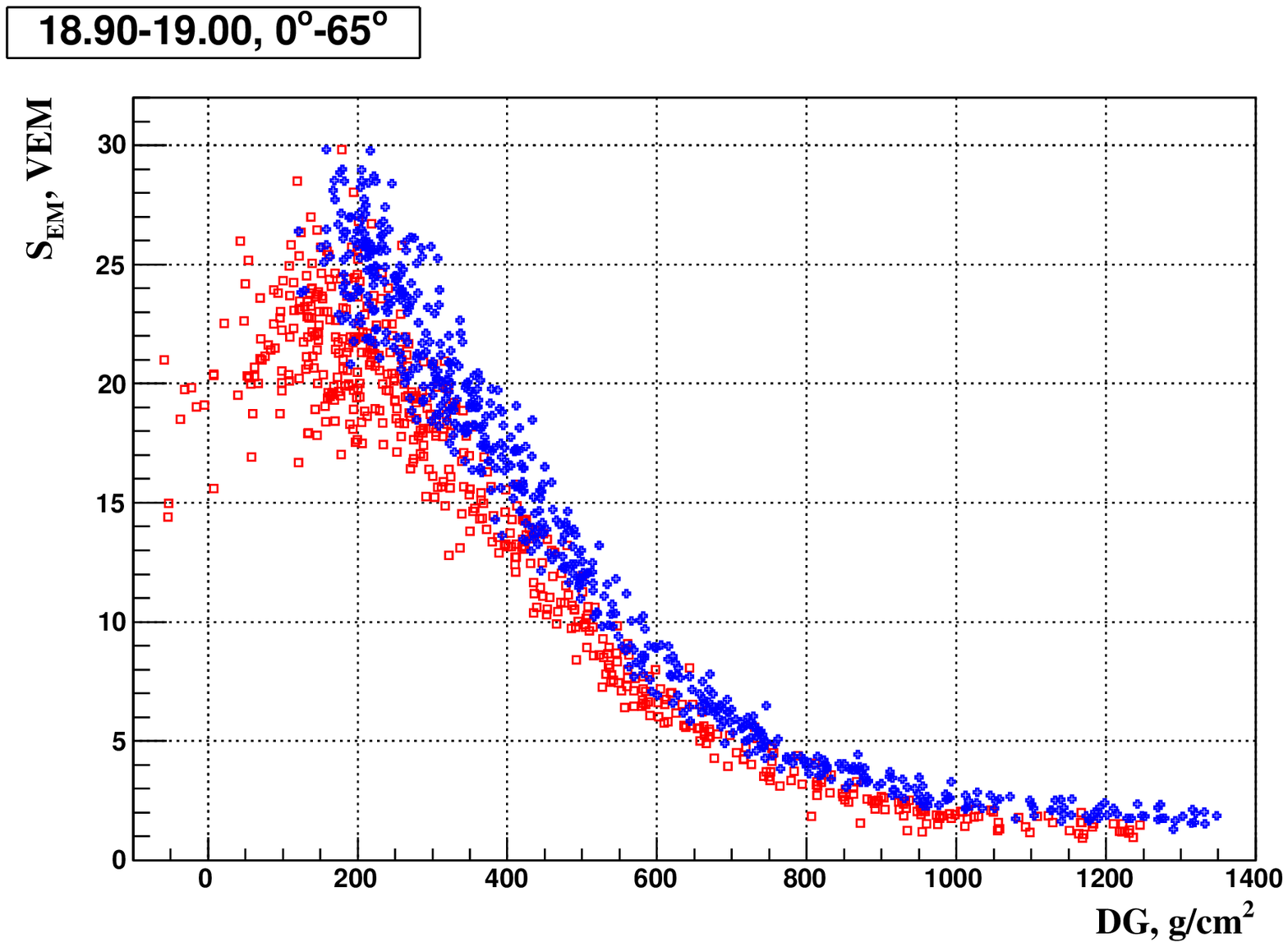}
\includegraphics[width=0.49\textwidth]{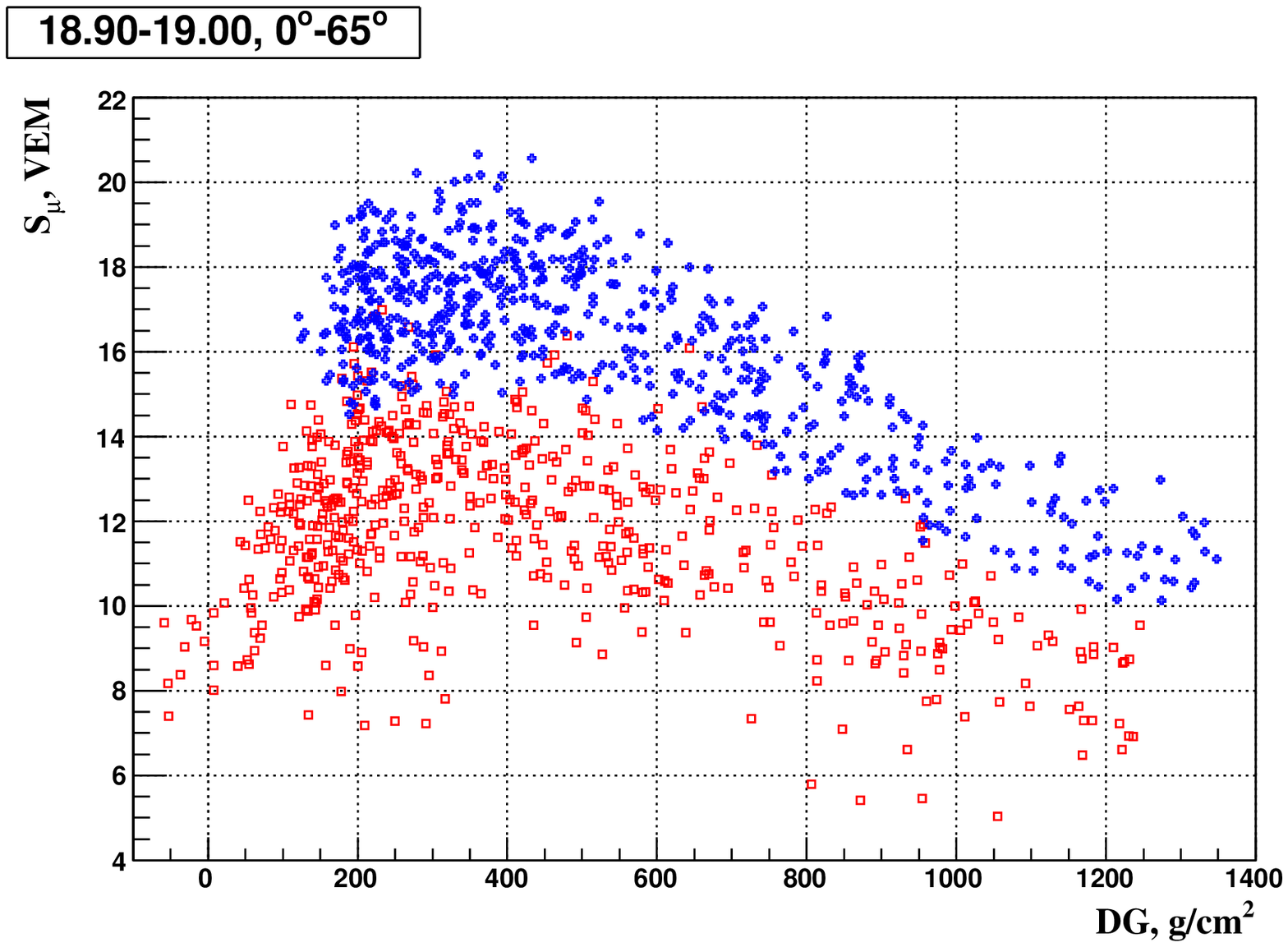}
\caption{EM and muon signals in showers from proton (red squares) and
  iron (blue crosses) in water Cherenkov tanks at 1000~m vs {\bf slant
  distance} from shower maximum to the ground $DG$ in $\logen18.9-19.0$
  energy bin and $\theta=0^\circ-65^\circ$ zenith angle range}
\label{DGSlant} 
\end{figure*}

\begin{figure*}
\includegraphics[width=0.49\textwidth]{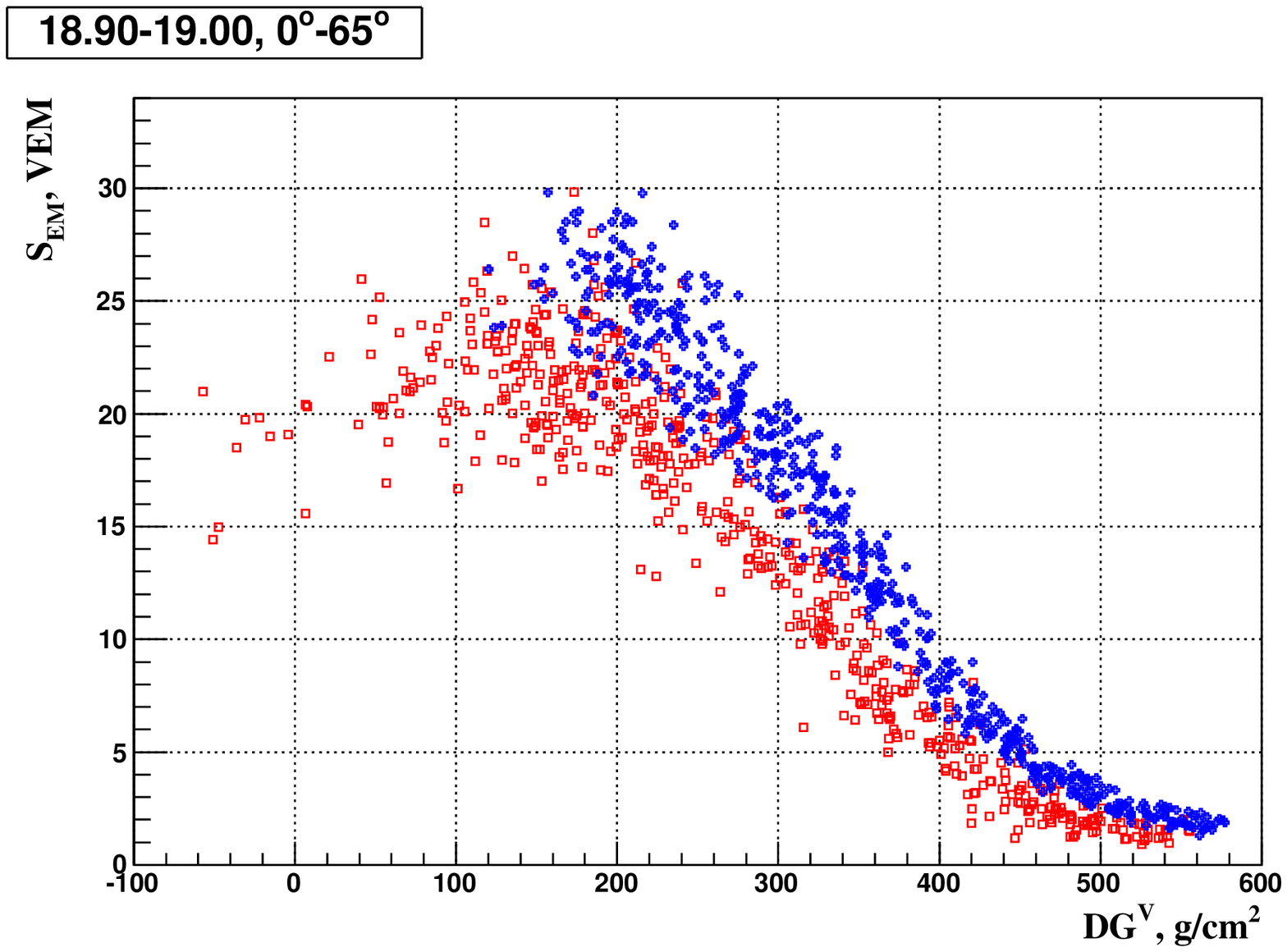}
\includegraphics[width=0.49\textwidth]{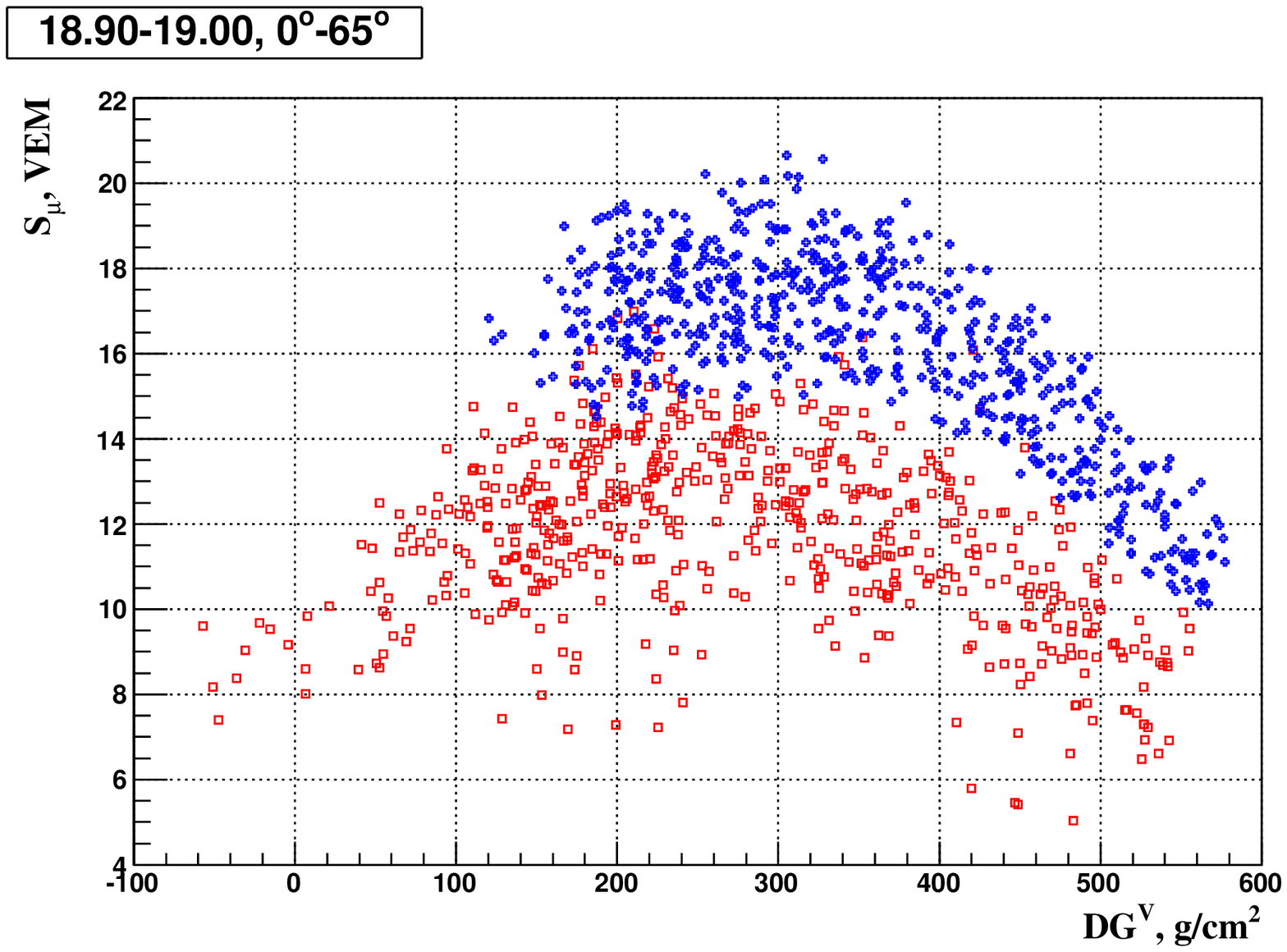}
\caption{EM and muon signals in showers from proton (red squares) and
  iron (blue crosses) in water Cherenkov tanks at 1000~m vs {\bf vertical
  distance} from shower maximum to the ground \dgv\ in $\logen18.9-19.0$
  energy bin and $\theta=0^\circ-65^\circ$ zenith angle range}
\label{DGVert}
\end{figure*}

\begin{figure*}
\includegraphics[width=0.49\textwidth]{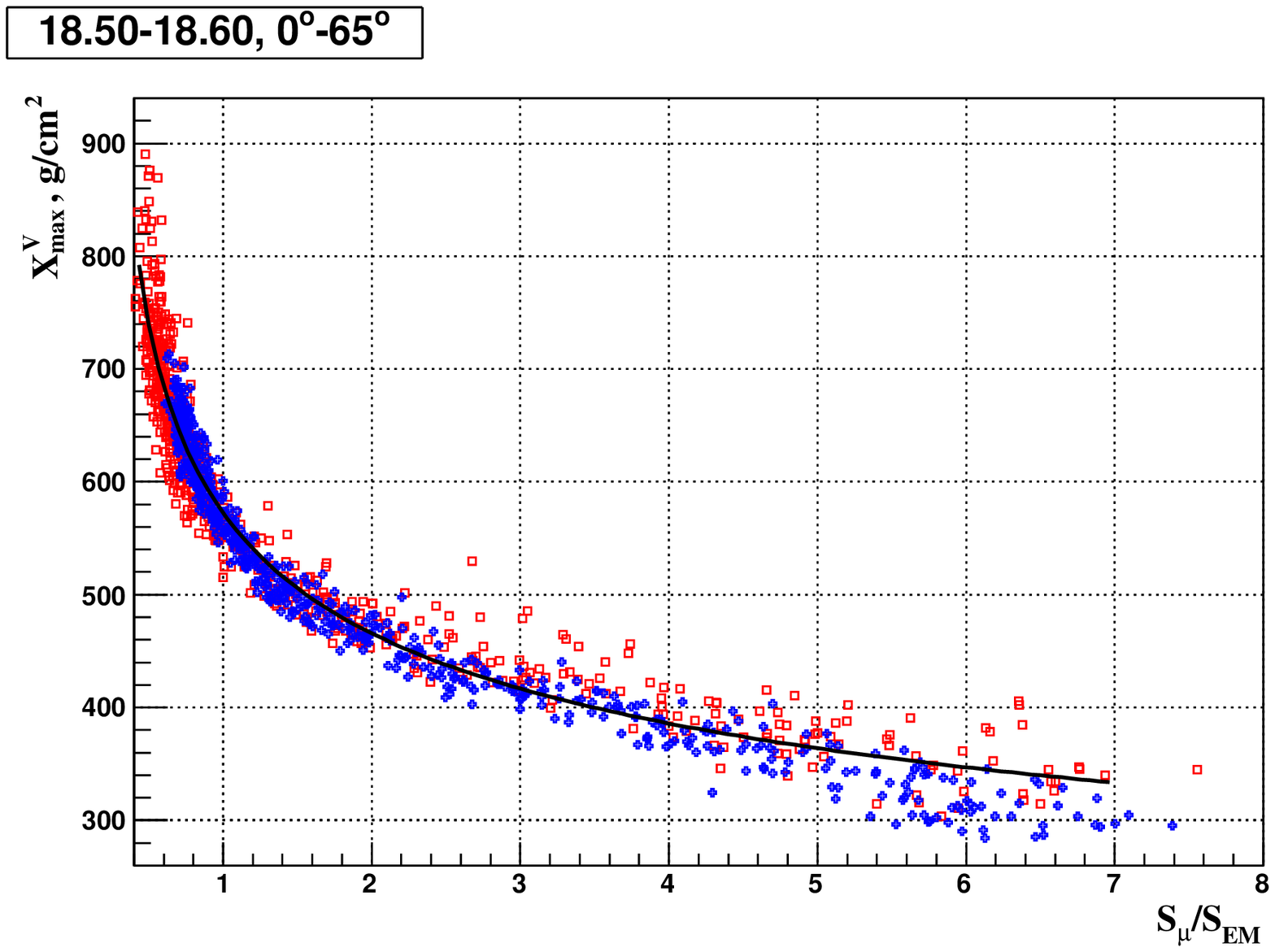}
\includegraphics[width=0.49\textwidth]{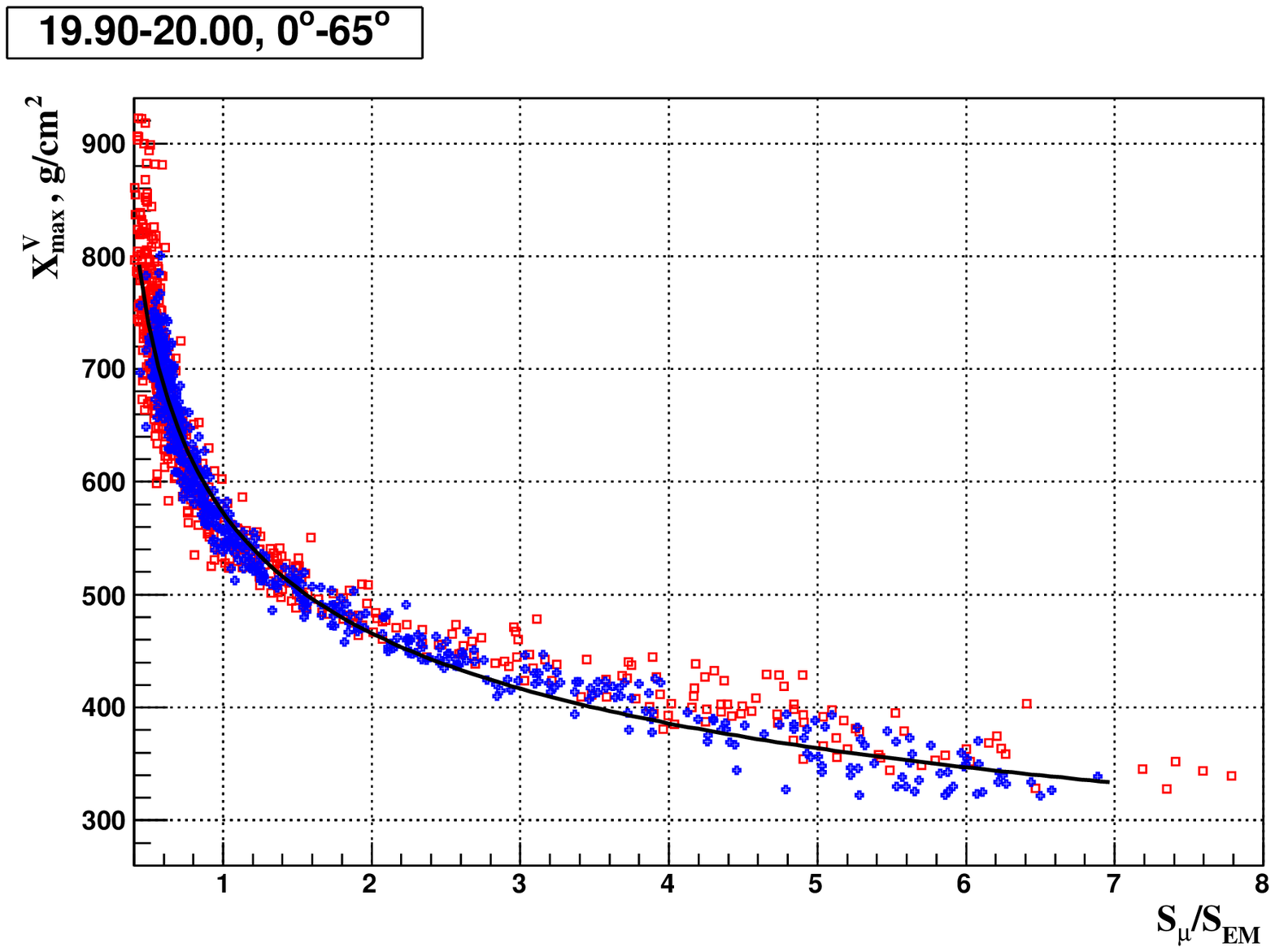}
\includegraphics[width=0.49\textwidth]{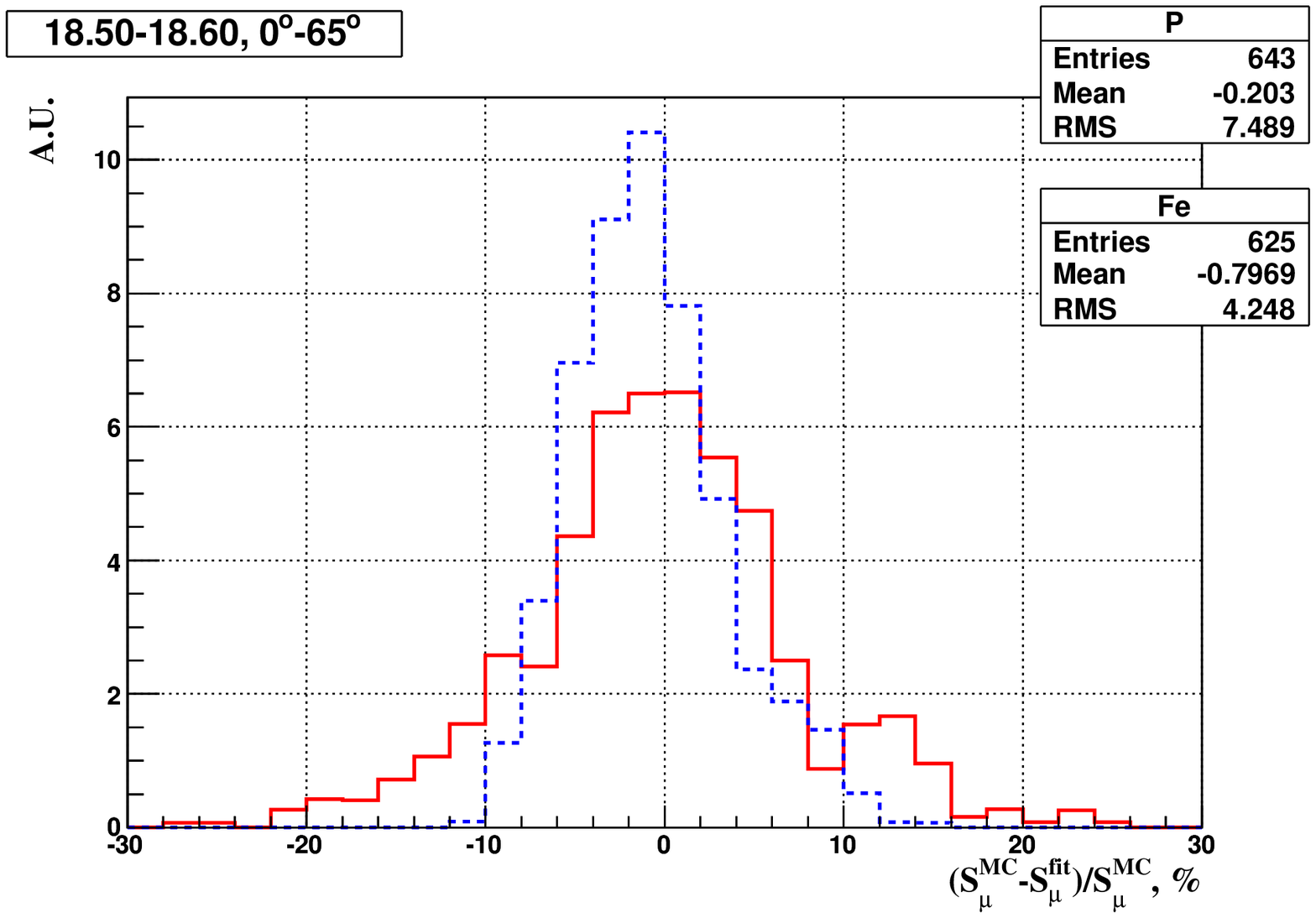}
\includegraphics[width=0.49\textwidth]{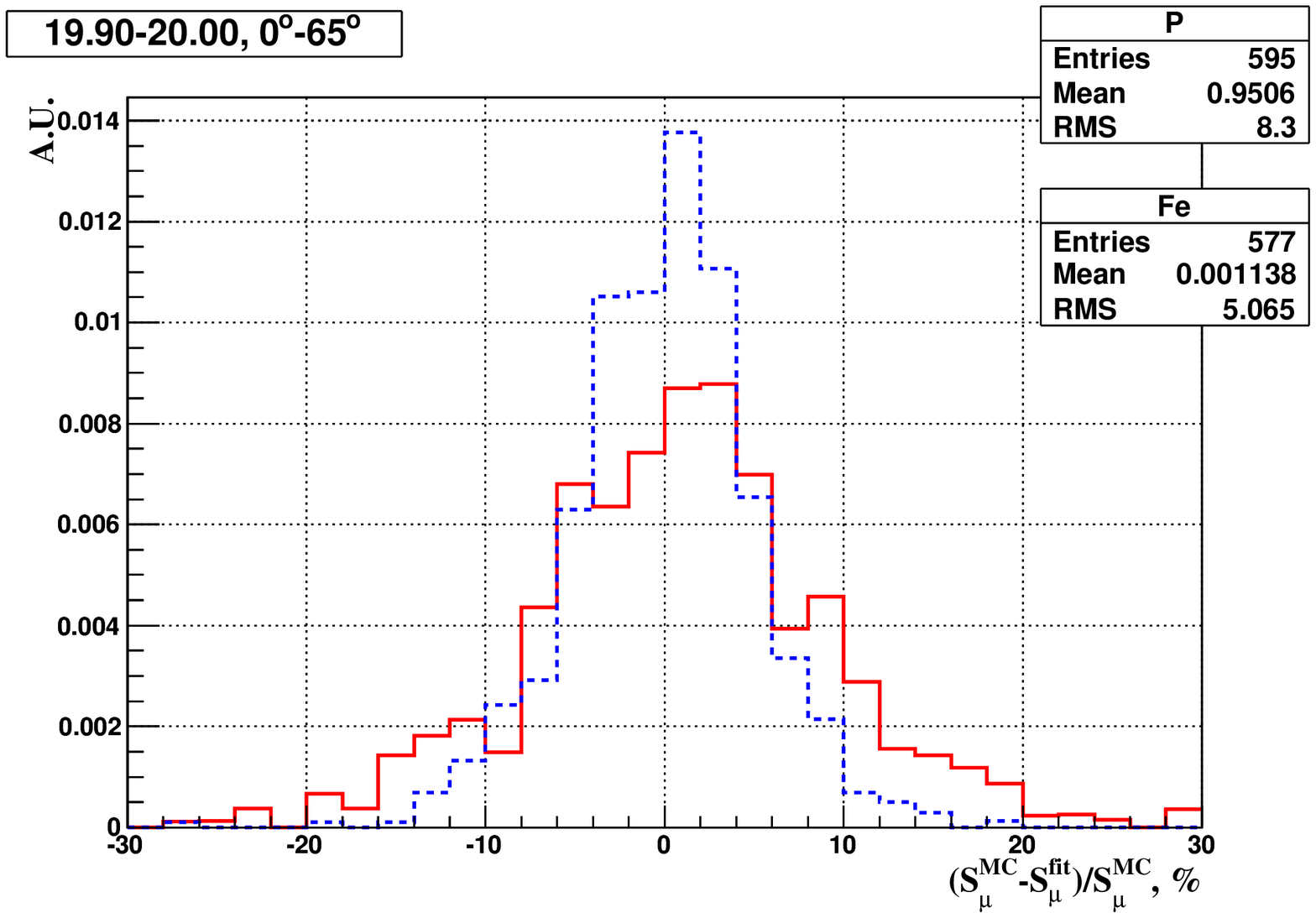}
\caption{Top: ratio of signals in water Cherenkov tanks \sigrat\ at
  1000~m vs vertical depth of shower maximum \xmaxv\ in two energy
  bins. Protons~---~red squares, iron~---~blue crosses. Black line is
  the fit in the form~(\ref{eq:fit}) with parameters specified in the text.
  Bottom: distributions of relative difference between MC simulated
  muon signals in Cherenkov water tanks $S_\mu^\mathrm{MC}$ and muon
  signals derived from the fit $S_\mu^\mathrm{fit}$ at
  1000~m. Protons~---~red solid line, iron~---~blue dashed line.}
\label{muemxmax}
\end{figure*}

\begin{figure}
\includegraphics[width=0.49\textwidth]{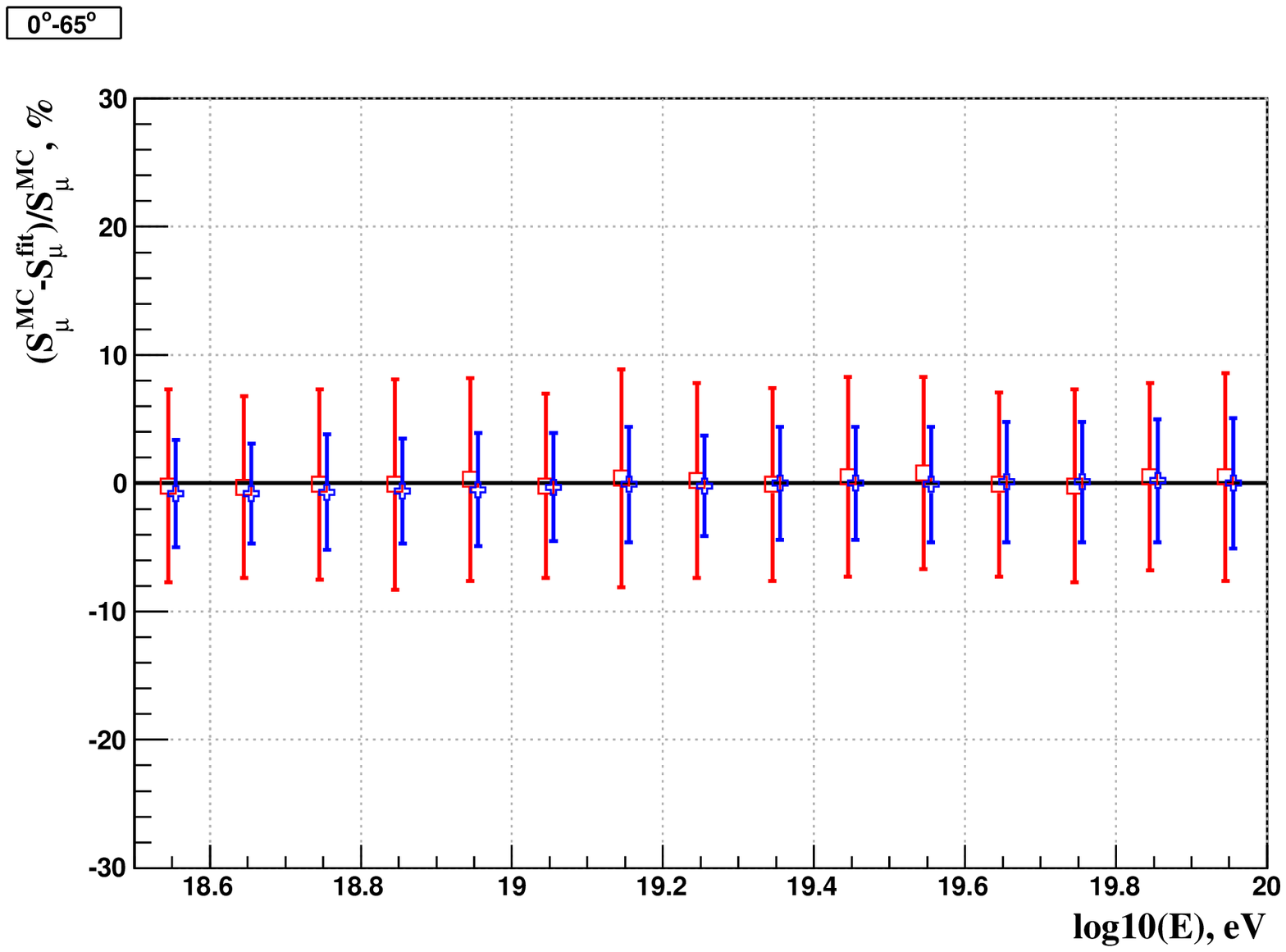}
\caption{Means and RMS of distributions of relative difference between
  MC simulated muon signals in Cherenkov water tanks
  $S_\mu^\mathrm{MC}$ and muon signals derived from the
  fit~(\ref{eq:fit}) $S_\mu^\mathrm{fit}$ at 1000~m (see also
  Fig.~\ref{muemxmax}), calculated with the unique set of parameters
  for all energy bins: $A=538$, $b=-0.25$, $a=-0.22$. Protons~---~red,
  iron~---~blue.}
\label{mudiff}
\end{figure}

\begin{figure*}
\includegraphics[width=0.49\textwidth]{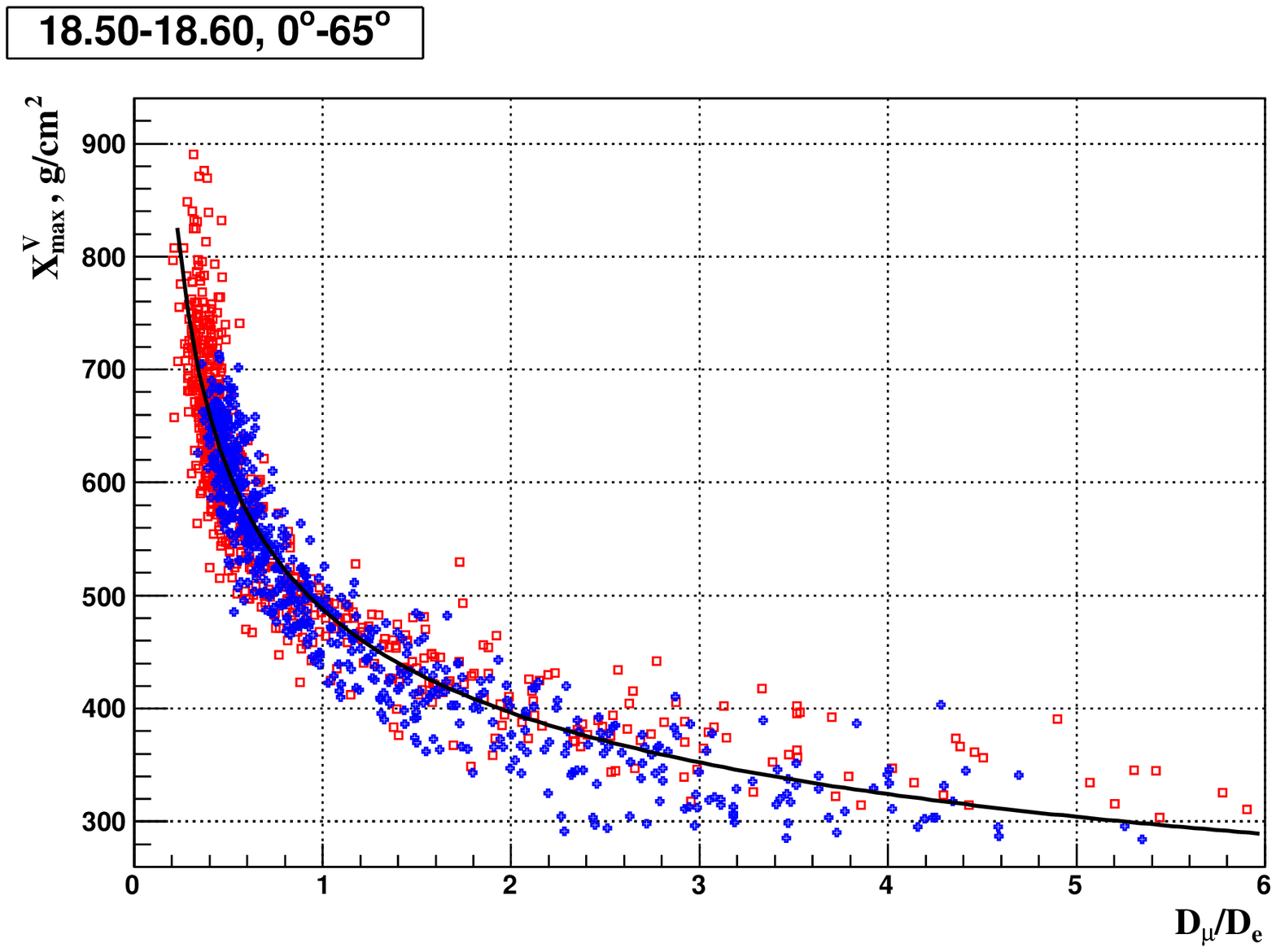}
\includegraphics[width=0.49\textwidth]{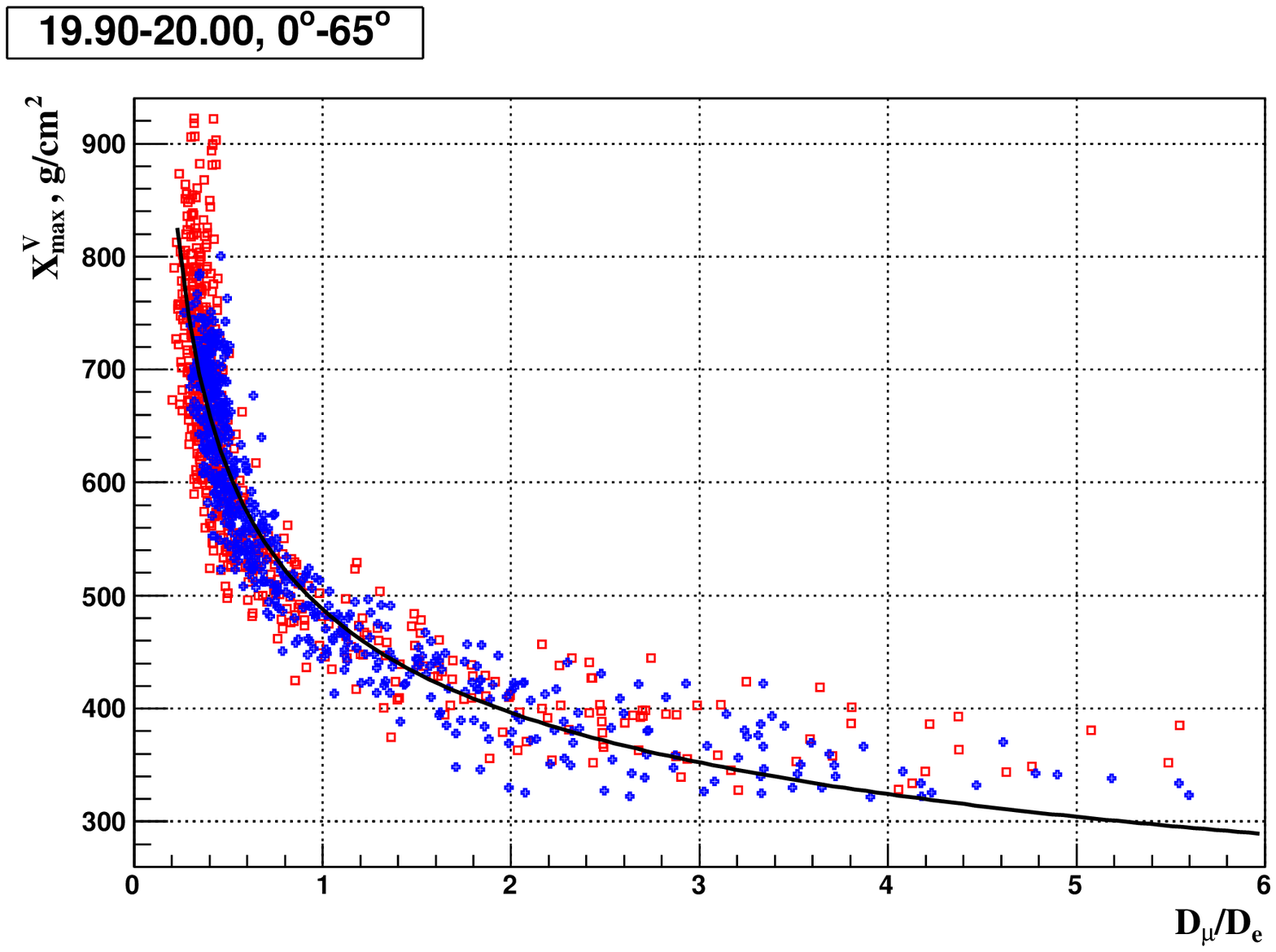}
\includegraphics[width=0.49\textwidth]{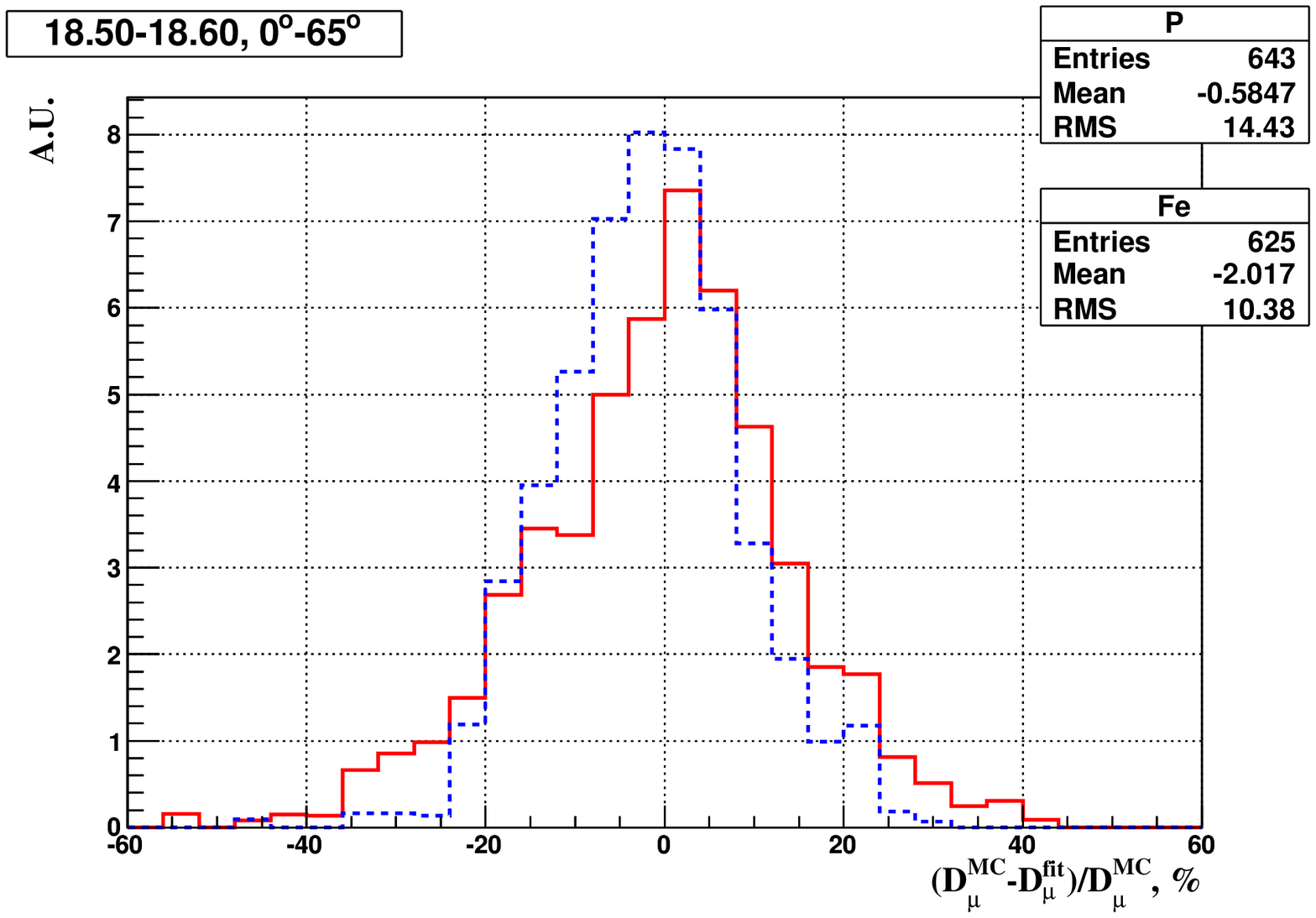}
\includegraphics[width=0.49\textwidth]{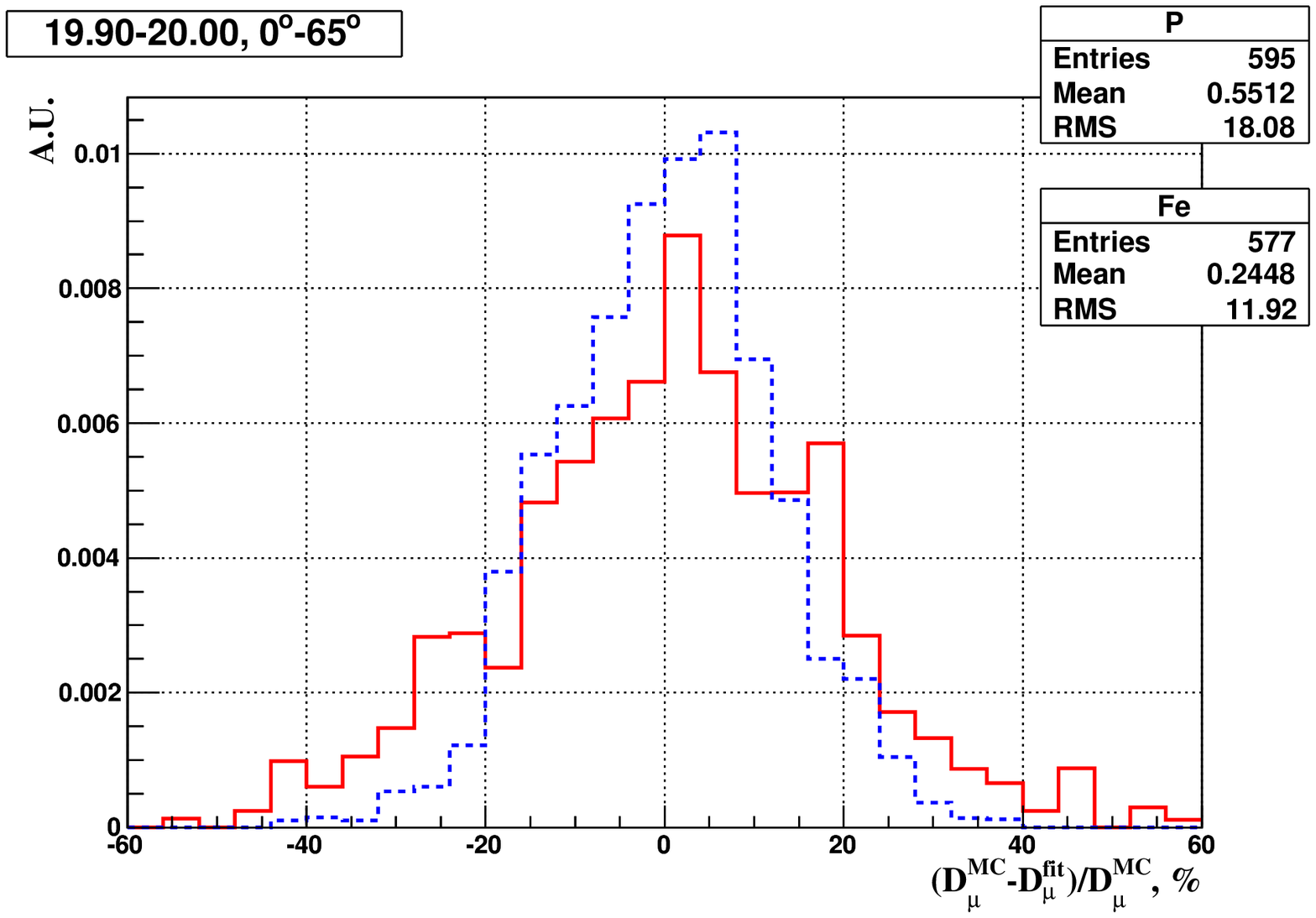}
\caption{Top: ratio of muon density to the electron one at 1000~m vs
  vertical depth of shower maximum \xmaxv\ in two energy
  bins. Protons~---~red squares, iron~---~blue crosses. Black line is
  the fit in the form~(\ref{eq:fit}) with parameters specified in the
  text. Bottom: distributions of relative difference between MC
  simulated muon density $D_\mu^\mathrm{MC}$ and muon density derived
  from the fit $D_\mu^\mathrm{fit}$ at 1000~m. Protons~---~red solid
  line, iron~---~blue dashed line.}
\label{Dmuemxmax}
\end{figure*}

\begin{figure}
\includegraphics[width=0.49\textwidth]{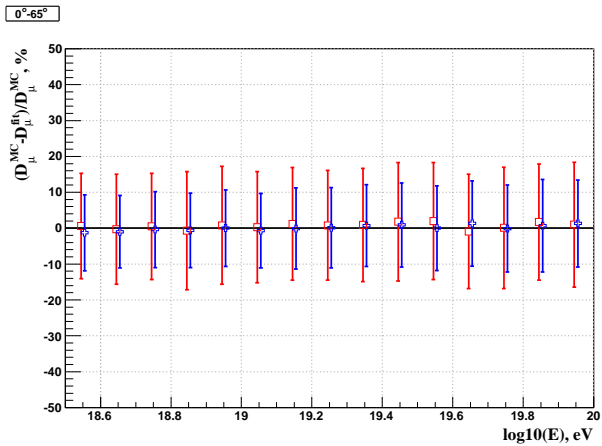}
\caption{Means and RMS of distributions of relative difference between
  MC simulated muon density $D_\mu^\mathrm{MC}$ and muon density
  derived from the fit $D_\mu^\mathrm{fit}$ at 1000~m (see also
  Fig.~\ref{Dmuemxmax}), calculated with the unique set of parameters
  for all energy bins $A=475$, $b=-0.28$, $a=-0.09$. Protons~---~red,
  iron~---~blue.}
\label{Dmudiff}
\end{figure}

\begin{figure*}
\includegraphics[width=0.49\textwidth]{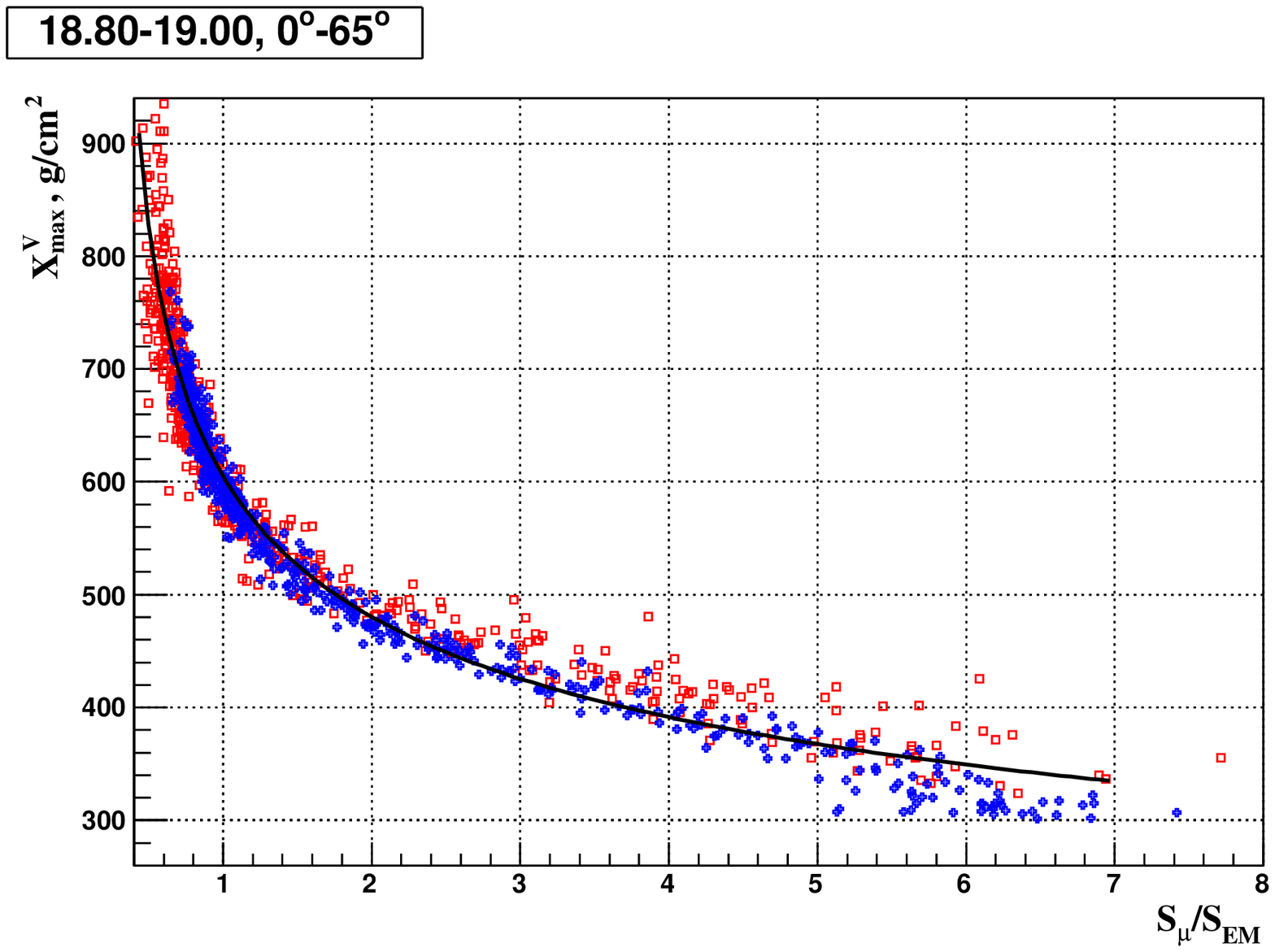}
\includegraphics[width=0.49\textwidth]{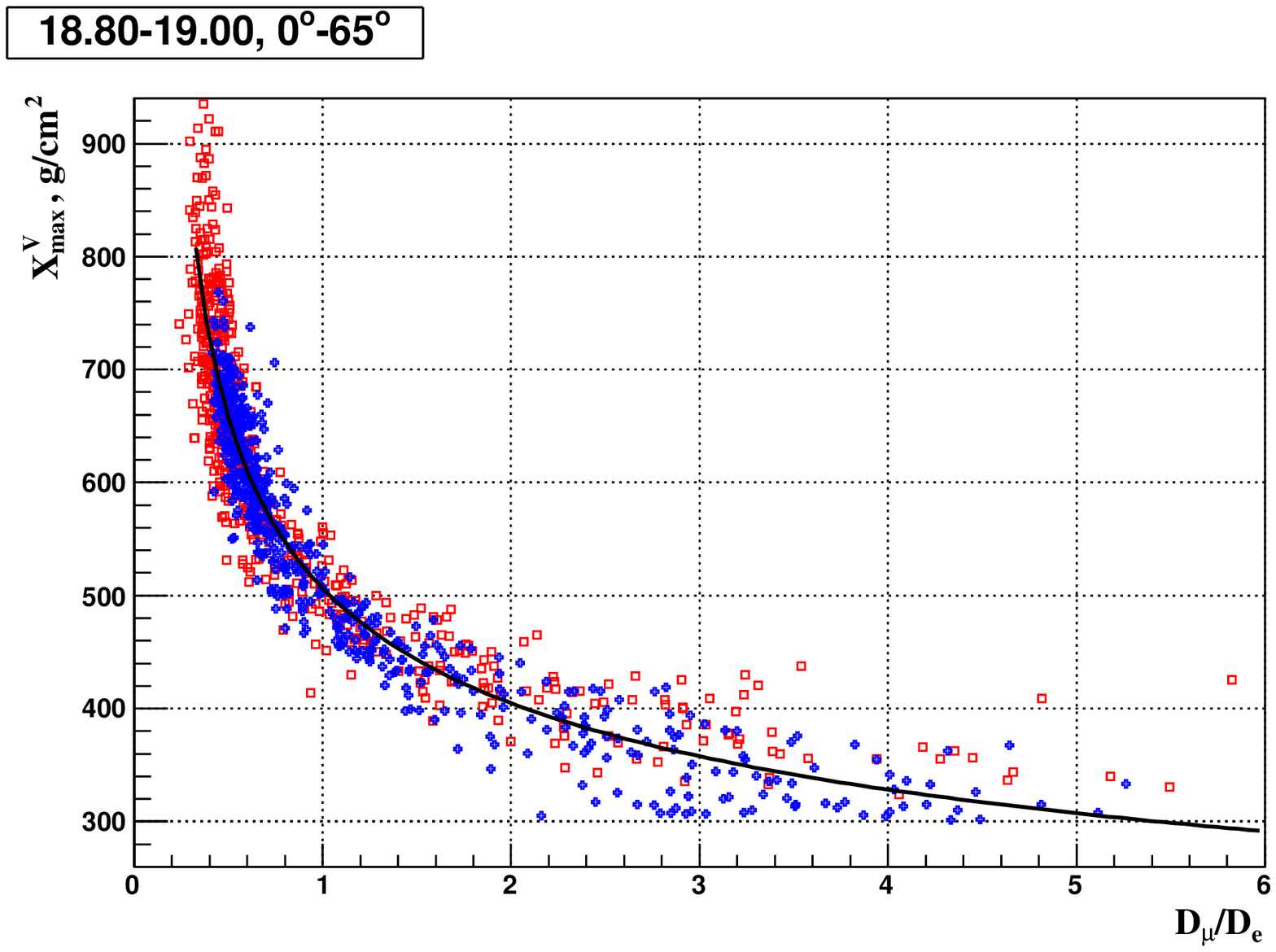}
\caption{Ratio of signals in water Cherenkov tanks \sigrat\ (left) and
  ratio of muon density to the electron one (right) at 1000~m vs
  vertical depth of shower maximum \xmaxv\ for EPOS~1.99 in
  $\logen18.8-19.0$ energy bin. Lines show the fit in the
  form~(\ref{eq:fit}), but coefficients certainly differ from
  those for QGSJET~II. The other designations are the same as in
  Fig.~\ref{muemxmax}.}
\label{epos_xmaxv}
\end{figure*}

\begin{figure*}
\includegraphics[width=0.49\textwidth]{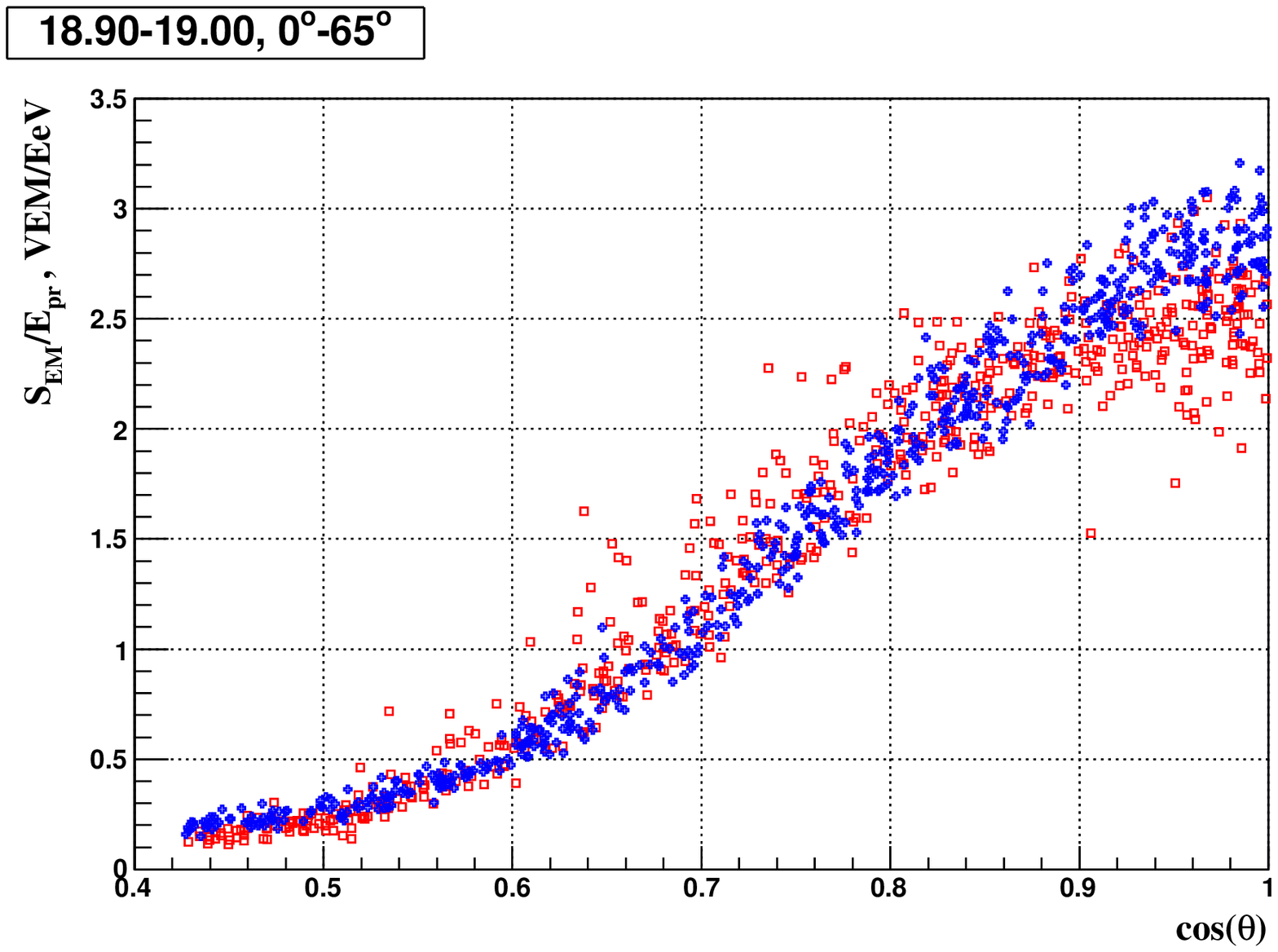}
\includegraphics[width=0.49\textwidth]{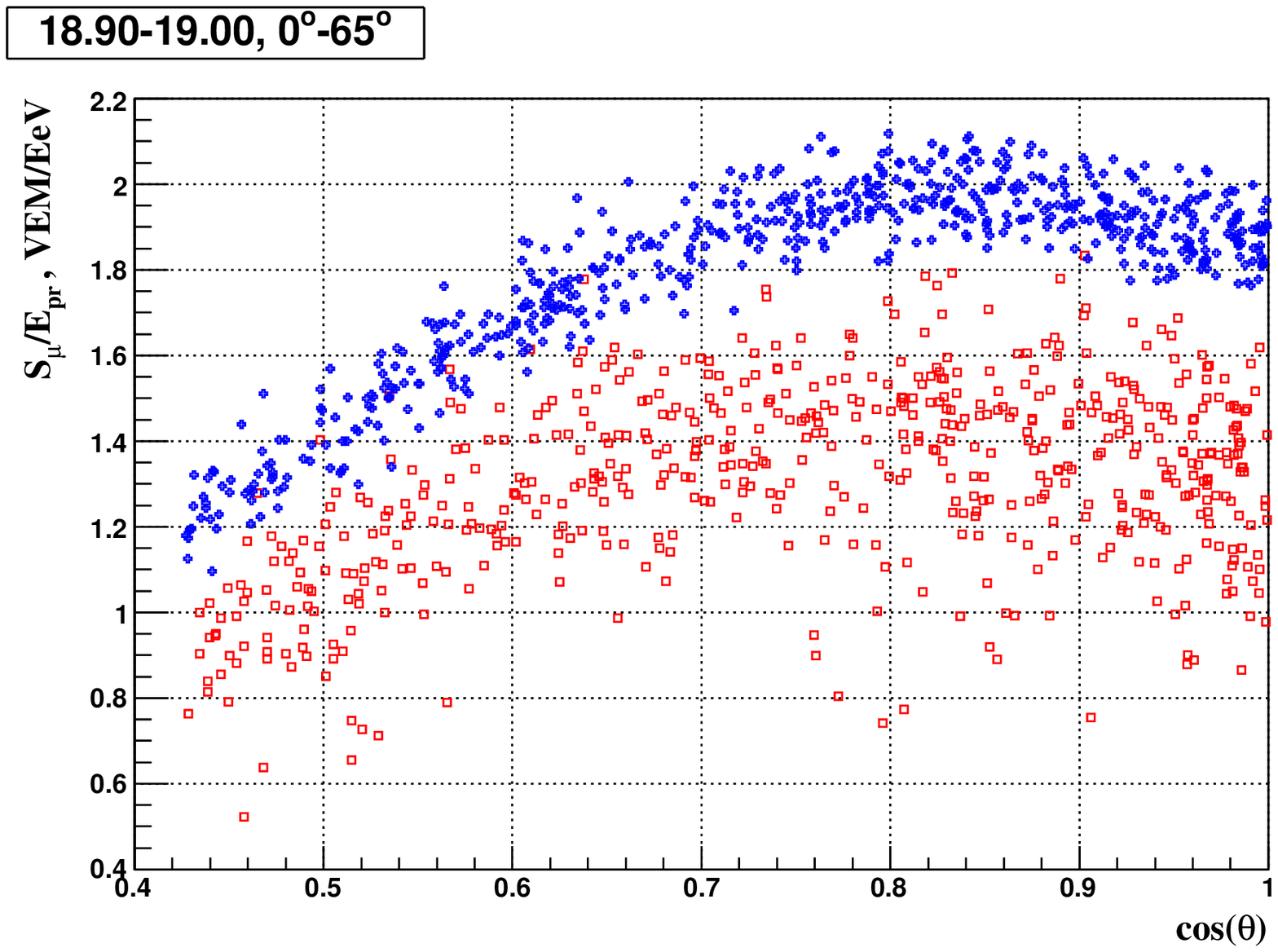}
\caption{EM and muon signals in showers from proton (red squares) and
  iron (blue crosses) in water Cherenkov tanks at 1000~m vs
  $\cos(\theta)$ in $\logen18.9-19.0$ energy bin and
  $\theta=0-65{}^\circ$ zenith angle range.}
\label{Scos}
\end{figure*}

\begin{figure*}
\includegraphics[width=0.49\textwidth]{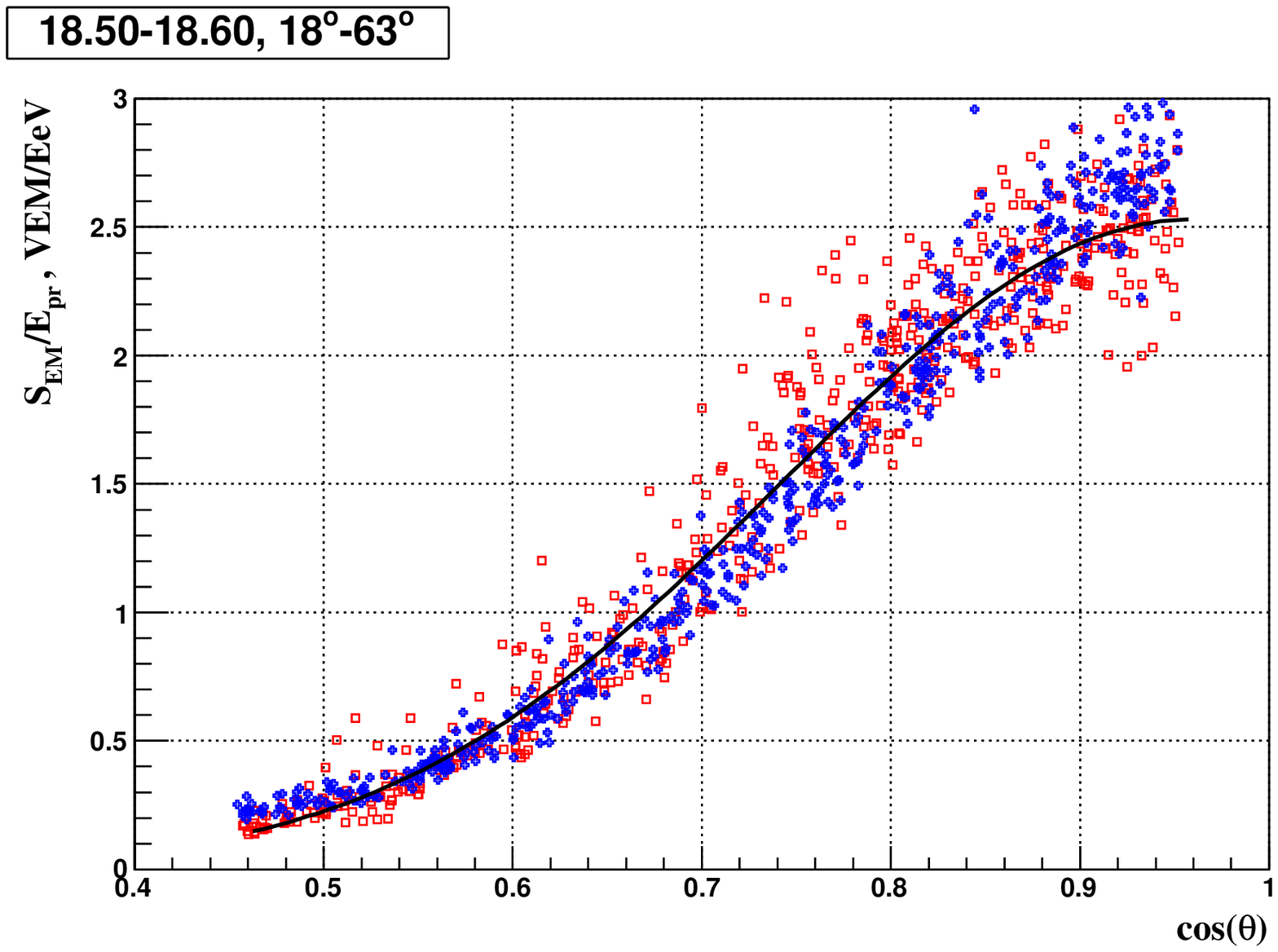}
\includegraphics[width=0.49\textwidth]{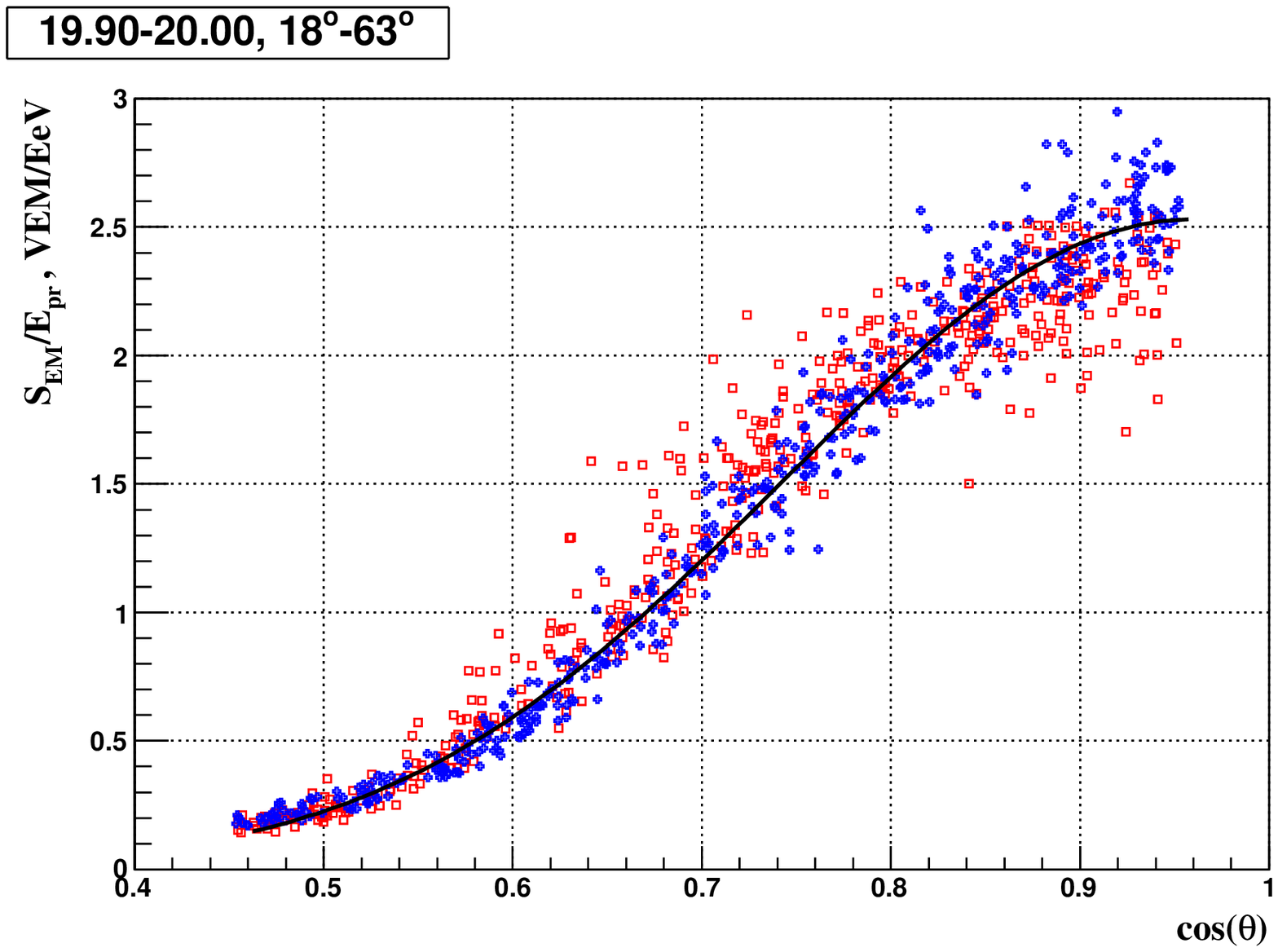}
\includegraphics[width=0.49\textwidth]{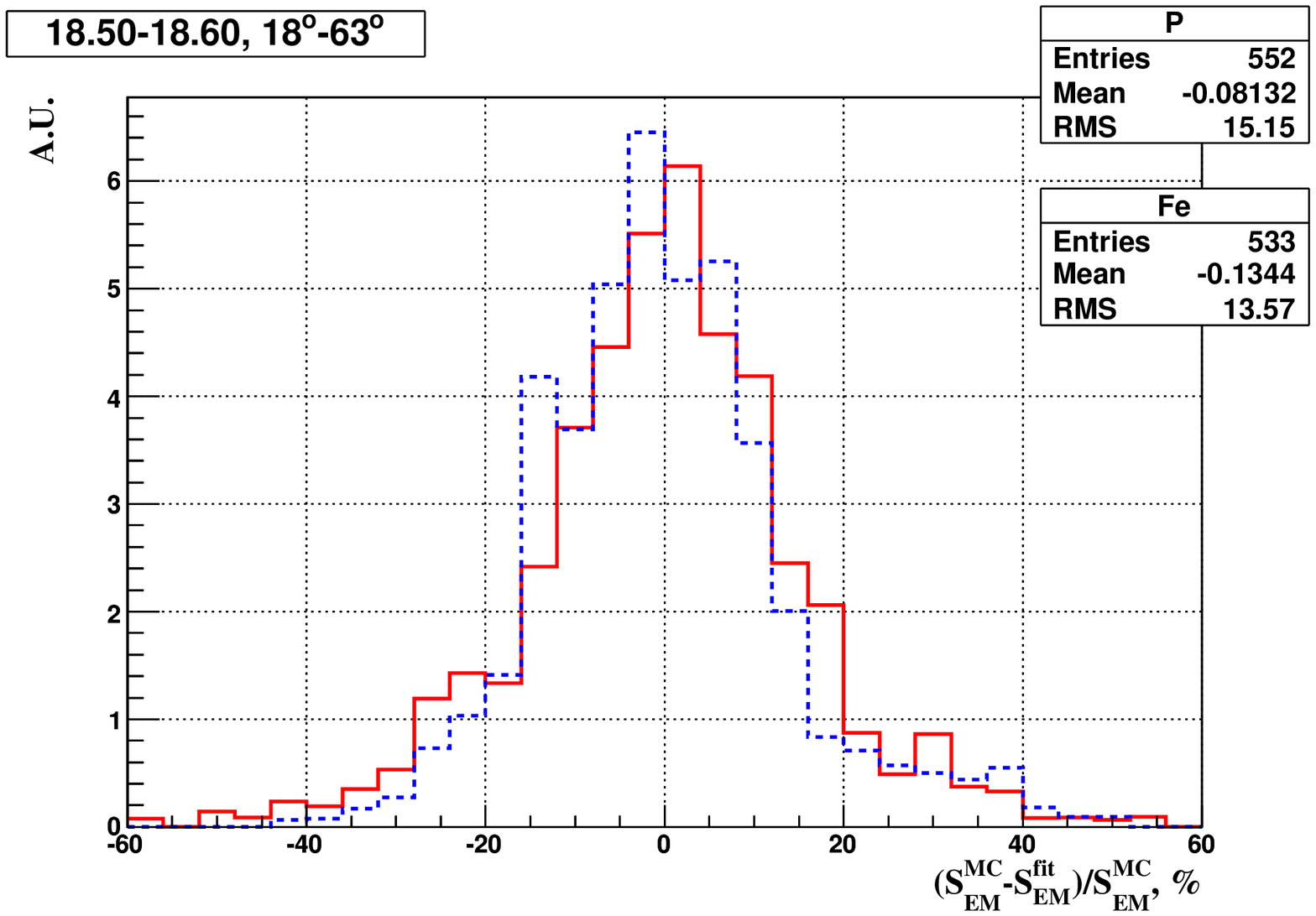}
\includegraphics[width=0.49\textwidth]{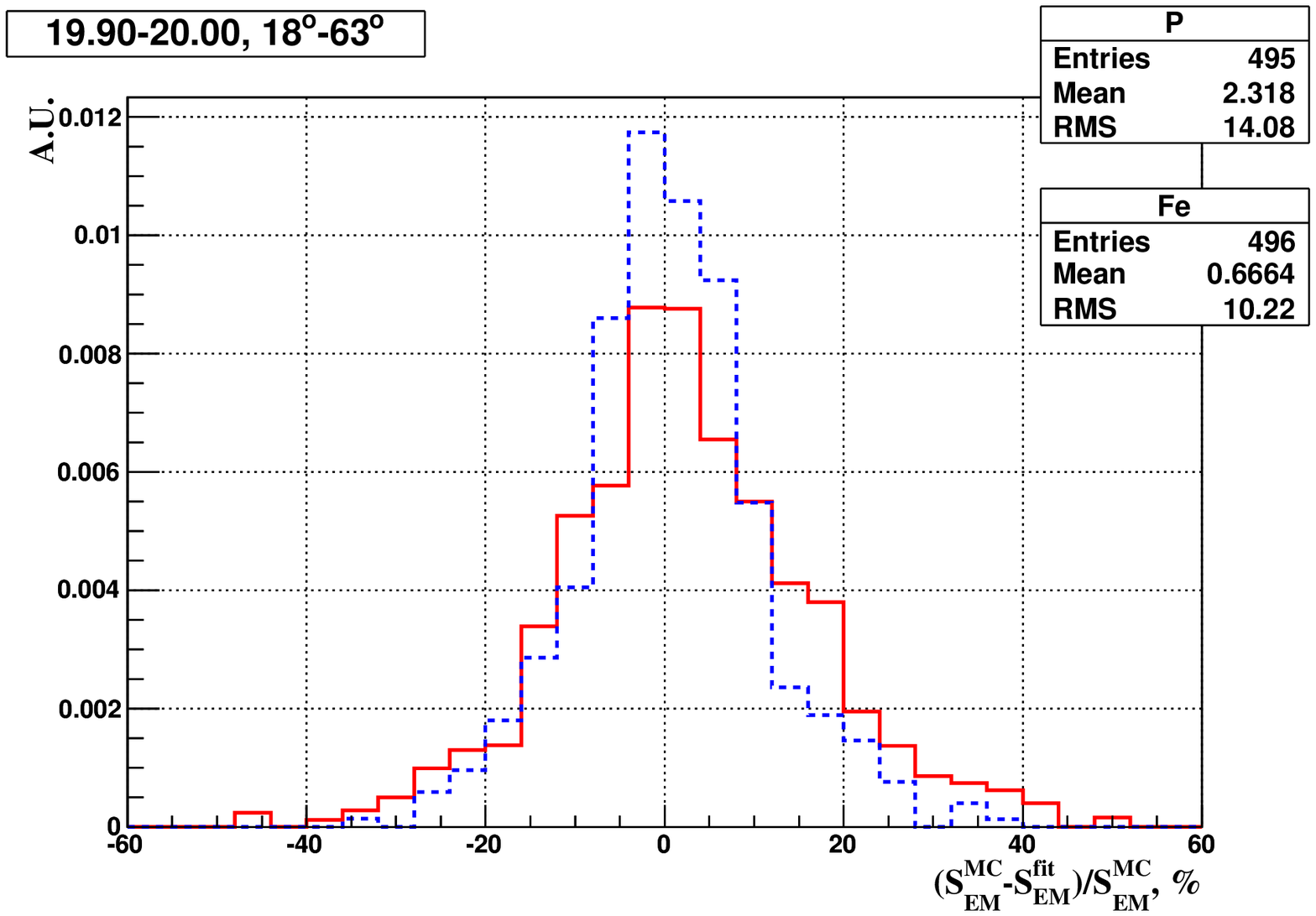}
\caption{Top: EM signals in water Cherenkov tanks at 1000~m vs
  $\cos(\theta)$ in two energy bins and $\theta=18-63{}^\circ$ zenith
  angle range. Black line is the fit in the form~(\ref{eq:semcos})
  with parameters specified in the text. Protons~---~red squares,
  iron~---~blue crosses. Bottom: distributions of relative difference
  between MC simulated EM signals in Cherenkov water tanks
  $\sem^\mathrm{MC}$ and muon signals derived from the fit
  $\sem^\mathrm{fit}$ at 1000~m. Protons~---~red solid line,
  iron~---~blue dashed line.}
\label{Sdiffcos}
\end{figure*}

\begin{figure}
\includegraphics[width=0.49\textwidth]{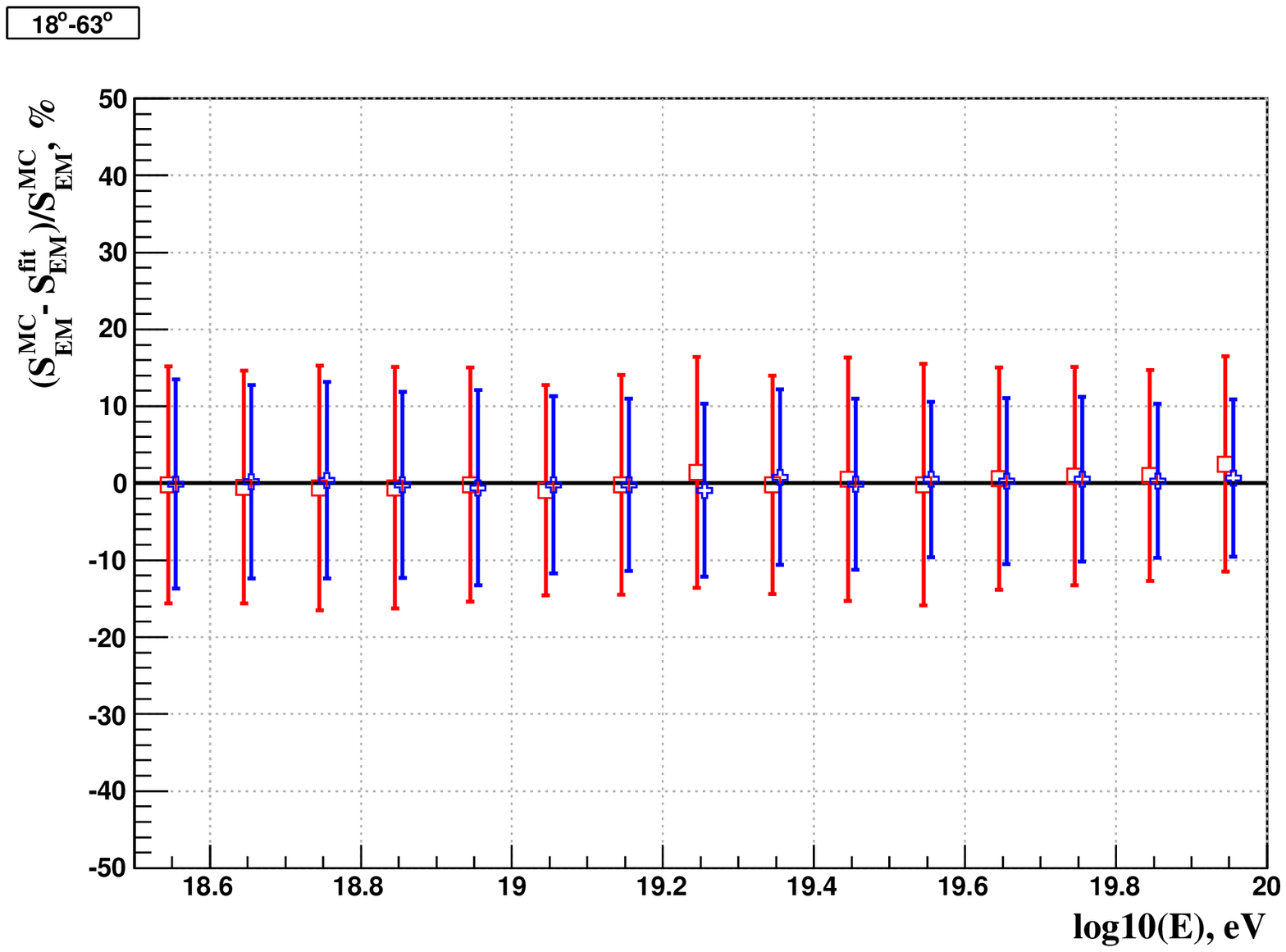} 
\caption{Means and RMS of distributions of relative difference between
  MC simulated EM signals in Cherenkov water tanks $\sem^\mathrm{MC}$
  and EM signals derived from the fit~(\ref{eq:semcos})
  $\sem^\mathrm{fit}$ at 1000~m (see also Fig.~\ref{Sdiffcos}),
  calculated with the unique set of parameters for all energy bins:
  $\sem^0=2.53$, $c_0=-3$, $c_1=0.96$, $\lambda=0.012$.
  Protons~---~red, iron~---~blue.}
\label{SemcosE}
\end{figure}

\begin{figure*}
\includegraphics[width=0.49\textwidth]{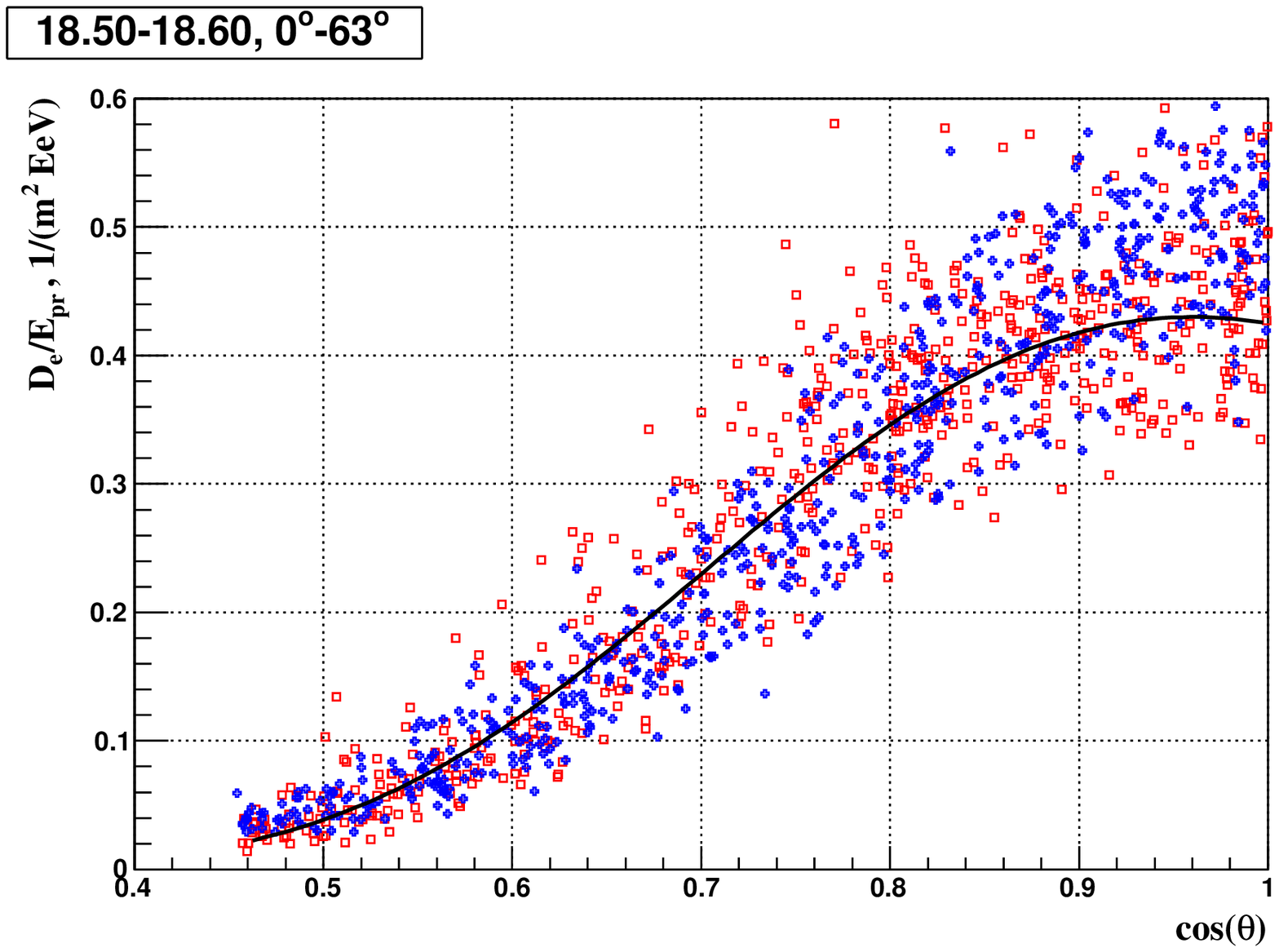}
\includegraphics[width=0.49\textwidth]{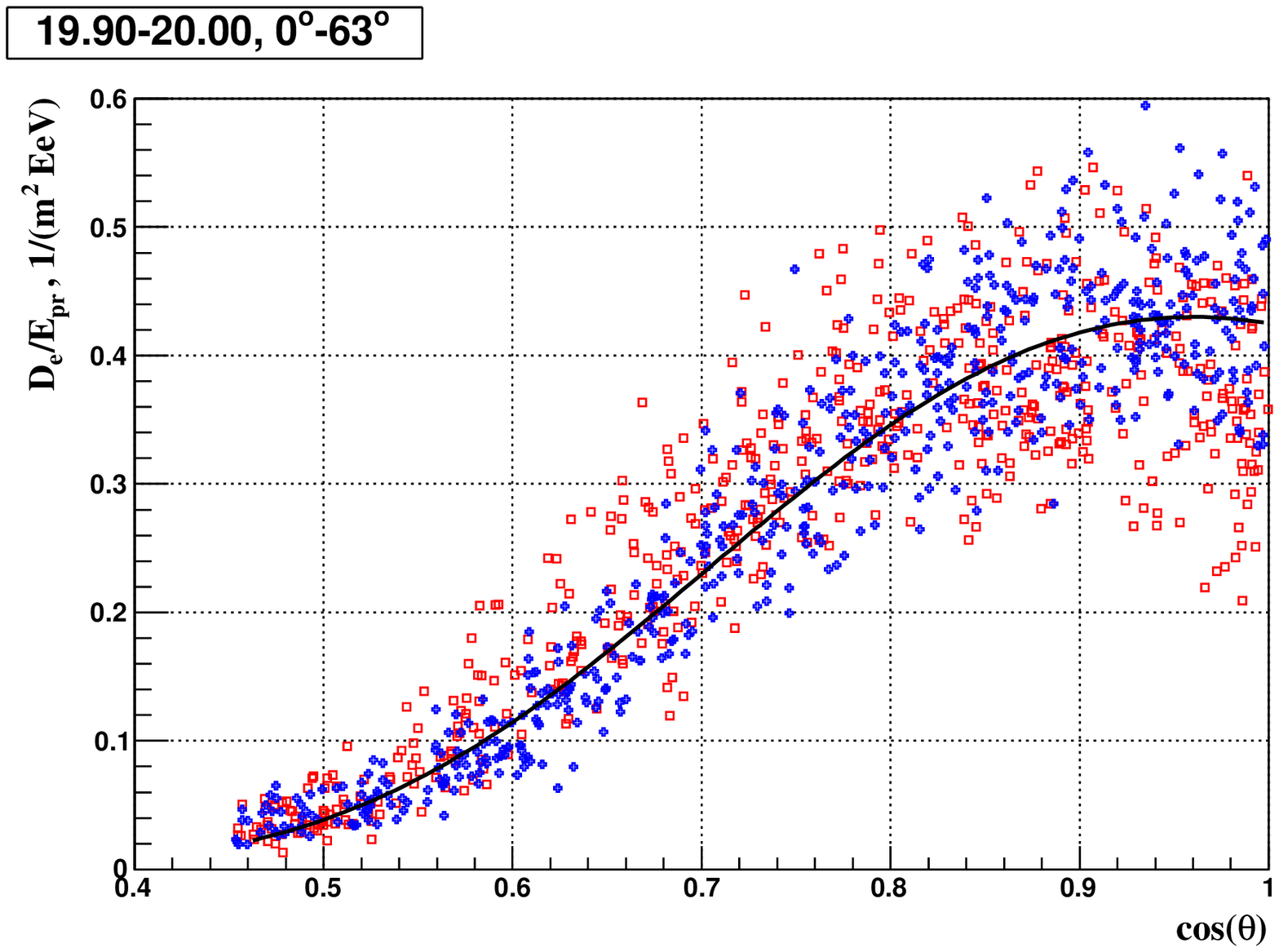}
\includegraphics[width=0.49\textwidth]{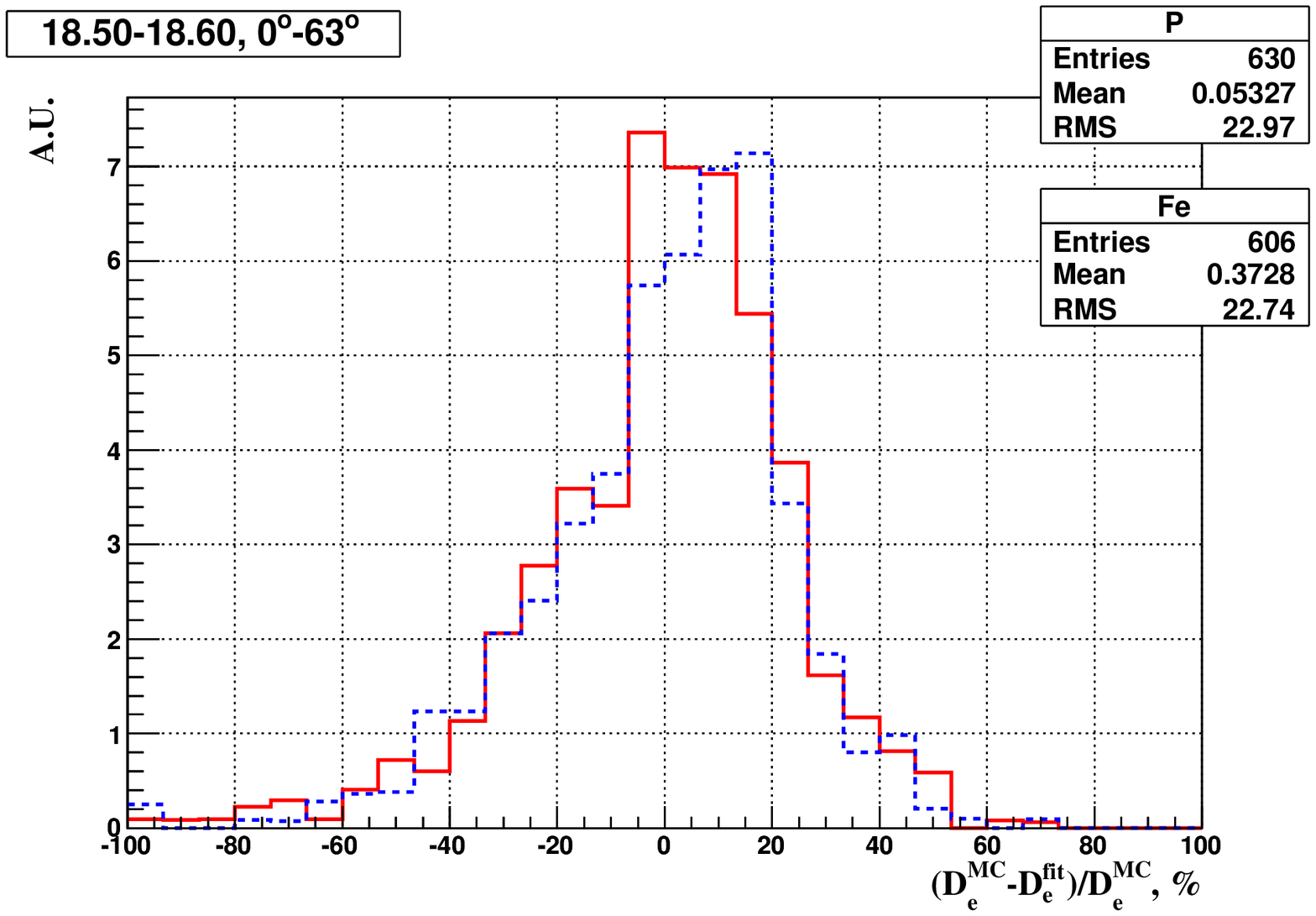}
\includegraphics[width=0.49\textwidth]{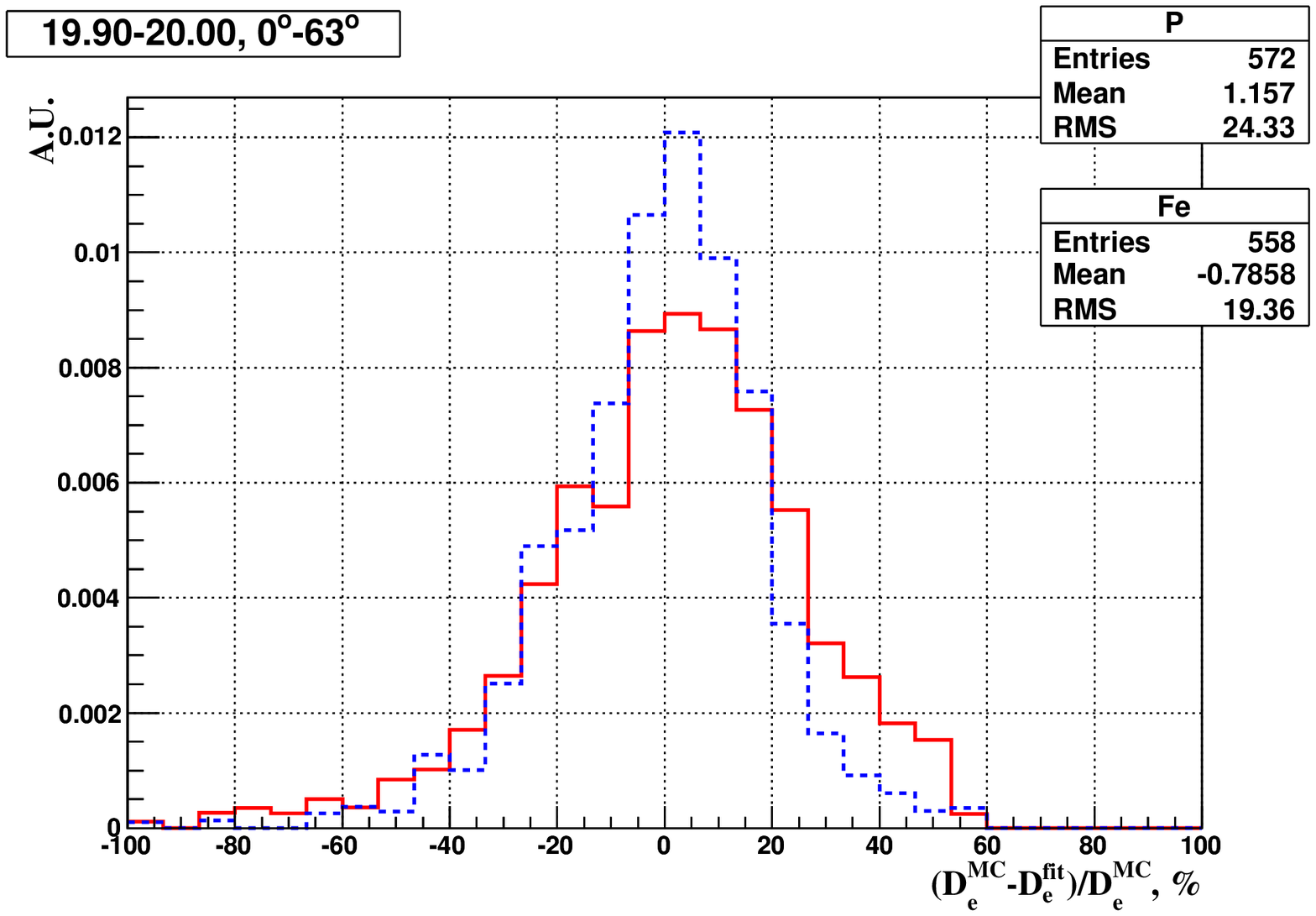}
\caption{Top: electron densities at 1000~m vs $\cos(\theta)$ in two
  energy bins. Black line is the fit in the form~(\ref{eq:semcos})
  with parameters specified in the text. Protons~---~red squares,
  iron~---~blue crosses. Bottom: distributions of relative difference
  between MC simulated electron densities $D_e^\mathrm{MC}$ and
  electron densities derived from the fit $D_e^\mathrm{fit}$ at
  1000~m. Protons~---~red solid line, iron~---~blue dashed line.}
\label{Ddiffcos}
\end{figure*}

\begin{figure}
\includegraphics[width=0.49\textwidth]{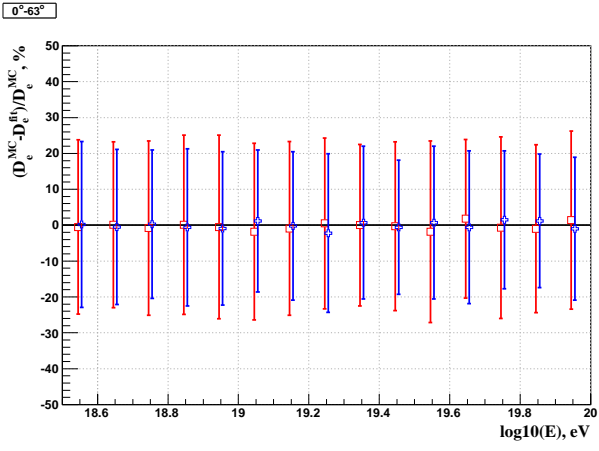}
\caption{Means and RMS of distributions of relative difference between
  MC simulated electron densities $D_e^\mathrm{MC}$ and electron
  densities derived from the fit~(\ref{eq:semcos}) $D_e^\mathrm{fit}$
  at 1000~m (see also Fig.~\ref{Ddiffcos}), calculated with the unique
  set of parameters for all energy bins: $D_e^0=0.43$, $c_0=0$,
  $c_1=0.96$, $\lambda=0.069$. Protons~---~red, iron~---~blue.}
\label{DcosE}
\end{figure}

\begin{figure*}
\includegraphics[width=0.49\textwidth]{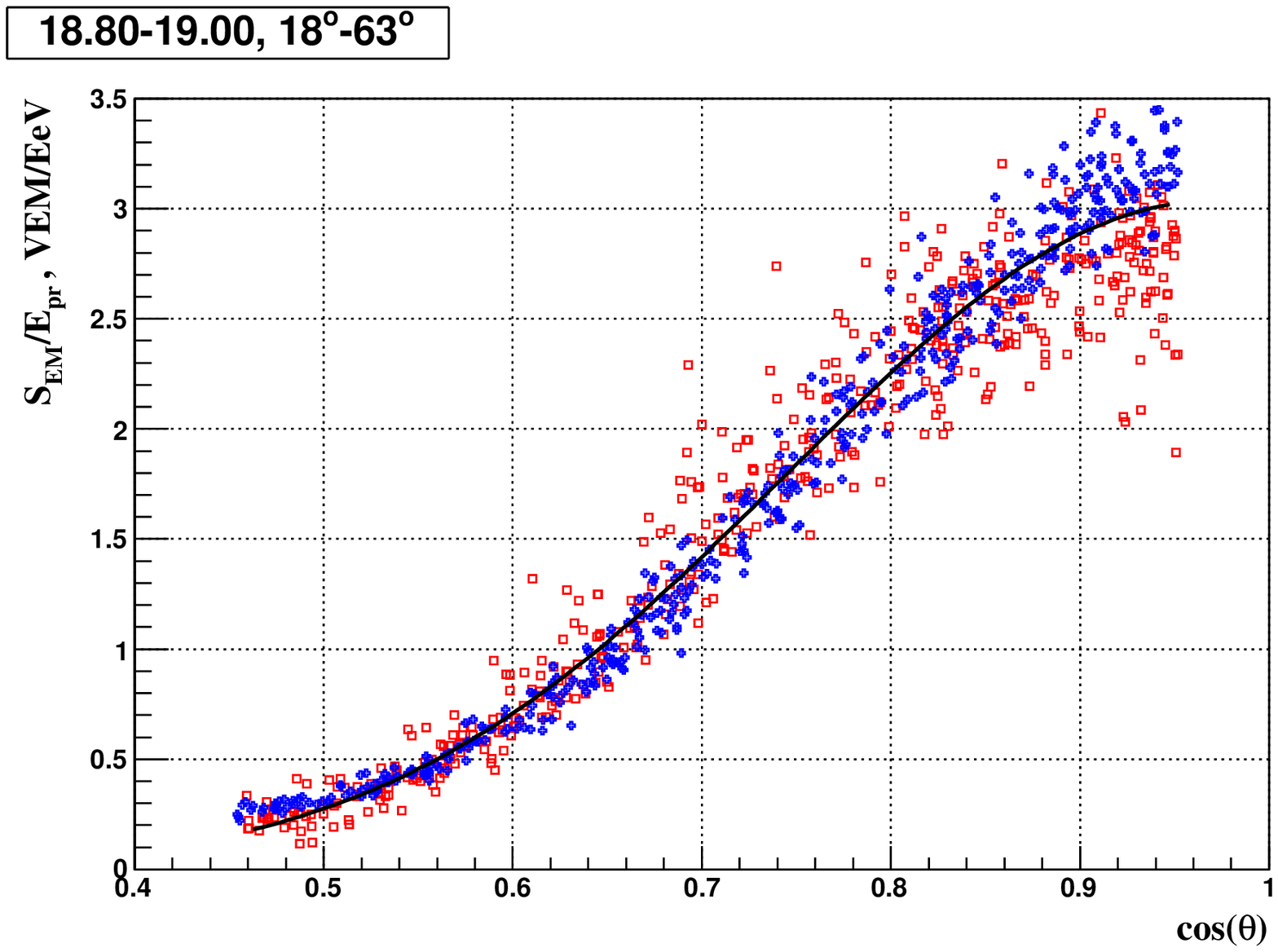}
\includegraphics[width=0.49\textwidth]{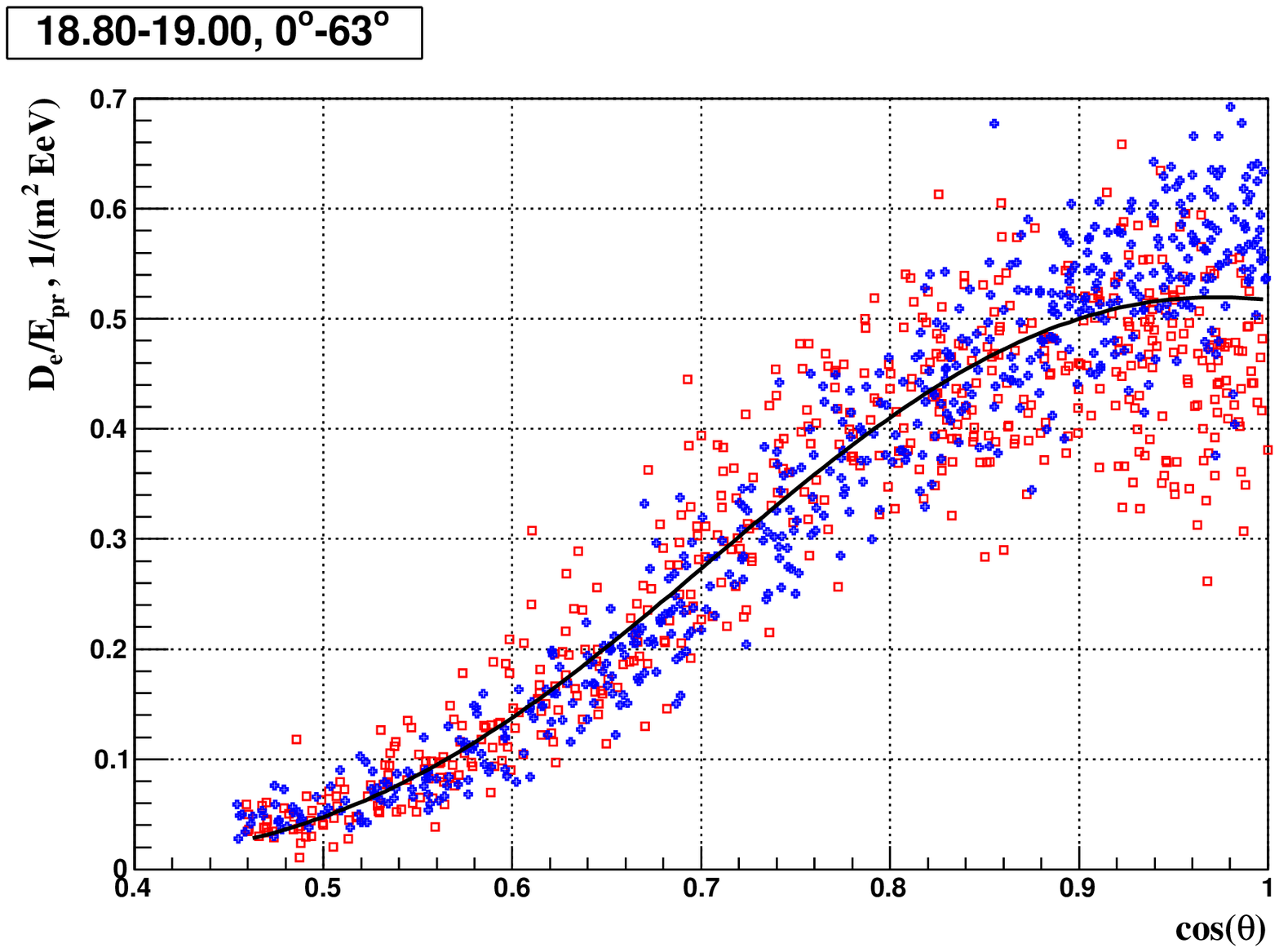}
\caption{EM signals in water Cherenkov tanks (left) and electron
  densities (right) at 1000~m vs $\cos(\theta)$ for EPOS~1.99 in
  $\logen18.8-19.0$ energy bin. Lines show the fits in the
  form~(\ref{eq:semcos}), but coefficients certainly differ
  from those for QGSJET~II. The other designations are the same as in
  Fig.~\ref{Sdiffcos}.}
\label{epos_cos}
\end{figure*}

\section*{Introduction}
Mass composition of ultra-high-energy cosmic rays (UHECR) can be
studied only indirectly with large EAS arrays. The contemporary
measurement of longitudinal and lateral shower characteristics in
hybrid
experiments~\cite{hires_aj01,PAO_proto_NIMA2004,TA_ISVHECRI2006}
provides the possibility to combine several primary mass sensitive EAS
parameters (such as depth of shower maximum and muon shower content)
to achieve the best primary particle mass
discrimination. Unfortunately, the lack of reliable information on
hadronic interaction properties at these energies causes large
uncertainties in the simulations of EAS characteristics and in turn
brings large uncertainties in mass composition analysis results (see
e.g. recent review~\cite{bluemer_crknee2009}). For most of the
interaction models the same experimental data show a lighter
composition from \xmax\ and a heavier (generally with mass above iron)
composition from muon shower
content~\cite{abuzayyad_mass2000,engel_icrc2007,ave_munum_2007,auger_massicrc2009}. Recently
proposed interaction model EPOS~\cite{werner_epos2006,epos_paris2009}
seems to be the first model providing quite consistent description of
longitudinal and lateral EAS profiles due to increased muon
production~\cite{epos_muprod2008}. EPOS application to Hires-MIA data
favors lightening of primary composition for energies above
$10^{18}$~eV well in agreement with Hires recent
results~\cite{socor,hires_p2009} (but compare with Yakutsk conclusions
on mixed composition obtained with EPOS~1.60 for energies above
$10^{19}$~eV~\cite{yakutsk_mu2008}) and can partly help to reduce the
25\% discrepancy between Auger fluorescence and surface detectors
energy scales~\cite{epos_muprod2008}. Nevertheless, this model has
still to be thoroughly tested against various EAS data in the wide
energy range and, what's more, for energies $E>10^{18}$, which are
well beyond the accelerators reach, there is no way to judge if
current EPOS (or any other model) properties correspond to actual
nature of hadronic interactions. The difference between the various
predictions of modern interaction models does not necessary cover all
possible range of accelerator data extrapolations (see
e.g.~\cite{bleve_31icrc}) which is so wide that allows even to treat
strong indications on primary mass increase for $E>10^{19}$~eV, which
can be derived from Auger data on \xmax\ and RMS of
\xmax\ distribution~\cite{socor}, within pure primary proton flux
hypothesis by modification of the total cross section or by
introduction of strong Feynman scaling
violation~\cite{alvarez_color_08,wibig_pp2009,wibig_pint2009,ulrich_cross2009}. Constraining
of such interaction parameters freedom is crucial for resolving of
UHECR mass composition problem, but it seems to be hardly achievable
on the basis of EAS studies alone and the data from the started LHC
are considered of fundamental importance.

In this paper, using universality property of EAS
development~\cite{giller_univers2005,nerling_univers2005,gora_univers2006,ave_icrc30_univers,lipari_univers2008,lafebre_univers2009},
we propose two simple, independent and accurate methods to determine
muon and EM shower contents and briefly discuss a possible way to test
and adjust interaction models in a primary mass independent way. We
also hope that the proposed EAS-universality-based correction of the
interaction models will allow to perform mass composition analysis
with the use muon EAS content in less interaction model dependent
manner.

The present study is performed making use of 28000 showers, generated
with CORSIKA~6.735~\cite{corsika} for $E^{-1}$ spectrum (and then
weighted to $E^{-3}$ spectrum) in the energy range $\logen18.5-20$
(with statistics of around 3000 showers for each primary in every of 3
energy bins $18.5-19.0$, $19.0-19.5$ and $19.5-20.0$) and uniformly
distributed in $\cos^2{\theta}$ in zenith angle interval
$\theta=0^\circ-65^\circ$ for
QGSJET~II~\cite{qgsjetii,qgsjetiia,qgsjetiib}/Fluka~\cite{fluka1,fluka2}
interaction models. The smaller set of around 10000 showers generated
with EPOS~1.99/Fluka was also used for verification of the presented
universality properties. EM component thinning was set to $10^{-6}$,
the observation level was at 870~\gsm, geomagnetic field was set to
Malarg$\ddot{\rm u}$e Auger site value.  All longitudinal shower
characteristics and charged particles density $D$, which is
effectively measured by  detectors using
scintillators~\cite{agasa}, were taken directly from CORSIKA particle
output files. The expected signal $S$ in Cherenkov Auger-like tanks
was calculated according to the sampling procedure described
in~\cite{billoir_sampl_2008,ave_munum_2007} with the use of the same
GEANT~4 lookup tables as in~\cite{ave_munum_2007}. Differently
from~\cite{ave_munum_2007} in this work the muon signal $S_\mu$
includes only signal from muons crossing the Cherenkov tank, while
signal from EM particles, originating from muon decays, is included in
the EM signal.

\section{Showers at the same vertical depth of maximum \xmaxv}

Of all aspects of universality of shower development we will be
interested only in EM and muon signals dependence on the distance of
shower maximum to the ground and on the zenith angle. Let's start from
Auger-like experimental setup and consider signal in water Cherenkov
tanks at 1000 meters from the shower core. In this case the common way
to express the universality of EM signal is to plot it against slant
distance to the ground (Fig.~\ref{DGSlant} and also Fig.~1
in~\cite{ave_icrc30_univers}), showing its quasi-independence from the
nature of the primary particle. The muon signal
functional dependence on slant distance to the ground $DG$ is also
very similar for both proton and iron, but there is a shift in the
normalization (Fig.~\ref{DGSlant}). Let us note, that since iron
showers reach \xmax\ earlier than showers from protons, comparison of
set of showers from p and Fe at equal $DG$ means comparison
between showers with different zenith angles, but at the same
development stage. Below, in this and in the next sections, we will
consider other two possible geometrical situations: showers having the
maximum at the same vertical depth \xmaxv\ and showers arriving at the
same zenith angles.

First, simply comparing shower characteristics dependence on vertical
distance to ground \dgv, one finds a very interesting property (see
Fig.~\ref{DGVert}). In this case the similarity of the functional
dependence of muon and EM signals on \dgv\ between p and Fe primaries
is preserved, but now also EM signal normalizations are
different. This happens because one compares showers, which have the
same vertical distance from \xmaxv\ to the ground, but proton showers
on average are more inclined than iron ones and therefore their EM
component attenuates more on the way from \xmax\ to the ground. The
ratio $S_\mathrm{EM}^\mathrm{Fe}/S_\mathrm{EM}^\mathrm{p}$ turns out
to be almost equal to the $S_\mu^\mathrm{Fe}/S_\mu^\mathrm{p}$ one and
this allows to state a new shower universality property: the ratio of
the muon signal to the EM one \sigrat\ is the same for all showers,
reaching the maximum at the same vertical depth \xmaxv, independently
on the primary particle nature, primary energy and incident zenith angle
(at least for the energy and angular ranges considered here). This
property is illustrated in Fig.~\ref{muemxmax}, where the dependence
of \sigrat\ on \xmaxv\ for p and Fe primaries is shown in two
different energy bins $\logen18.5-18.6$ and $19.9-20.0$. The
functional dependence between \xmaxv\ and \sigrat\ turns out to be
very simple and quasi-universal for all energies and primaries. The
function in the form
\begin{equation}
\label{eq:fit}
\xmaxv=A(S_\mu/S_\mathrm{EM}+a)^b 
\end{equation}
has been used to fit the data in 15 energy bins $\Delta\log10(E)$[eV]=0.1
and the fit parameters have been found to be stable across the entire
energy range. Using the functional dependence of \xmaxv\ on (\sigrat)
and $S_\mathrm{tot}=S_\mathrm{EM}+S_\mu$ one easily gets the equation,
which allows to obtain the muon signal from shower vertical depth and
total signal in water Cherenkov tanks:
\begin{equation}
\label{eq:mufit}
S_\mu^\mathrm{fit}=\frac{S_\mathrm{tot}}{1+1/(\left(\xmaxv/A\right)^{1/b}-a)}.
\end{equation}
We calculated the difference between the Monte-Carlo (MC) simulated
muon signal $S_\mu^\mathrm{MC}$ and the muon signal obtained from the
fit $S_\mu^\mathrm{fit}$, examples of its distribution are shown in
Fig.~\ref{muemxmax}. In Fig~\ref{mudiff} we plot behaviour of mean and
RMS values of these distributions for various energy bins, obtained
with the unique set of fit parameters $A=538$, $b=-0.25$ and $a=-0.22$
which represent their averages over 15 $\Delta\log10(E)$[eV]=0.1
energy bins. It is seen that the estimates of muon signals are
unbiased with less than 1\% deviation of the mean reconstructed muon
signal from the MC one for all primaries and the RMS values are small:
8\% for protons and around 5\% for oxygen and iron (though we don't
show results for oxygen which are always between p and Fe, we used
oxygen showers together with proton and iron ones to perform
fits). Certainly, the application of specific coefficients for each
energy bin, or narrowing of zenith angle interval, or using of more
sophisticated fit functions can possibly slightly improve the
performance of the method, but its simple and universal form already
works very well. The described universal dependence of \sigrat\ on
\xmaxv\ holds true in the wide interval of distances (at the least
from 200 to 1500 meters~\cite{ya_mu2009}) from the core, though for
distances closer to the core the function in the form~(\ref{eq:fit})
does not describe accurately the data in the entire angular range
$0^\circ-65^\circ$ and splitting the angular range in two regions or
applying more complex parametrization may be needed.

The same universality principle holds for the density of charged
particles. We have performed the reconstruction of muon densities
using the dependence of the ratio (muon density $D_\mu$)/(electron
density $D_\mathrm{e}$) on \xmaxv\ as in equation~(\ref{eq:fit})
with a unique set of parameters for all energy bins $A=475$,
$b=-0.28$, $a=-0.09$ (see Figs.~\ref{Dmuemxmax},\,\ref{Dmudiff}). It
is seen that when muons and electrons equally contribute to the
detector signal, the shower fluctuations play a more important role
and the accuracy of parametrization is only within 15\%, though the
estimate is still unbiased.

In this context we would like to emphasize that since our study does
not take into account any experimental error, we use the
parametrizations only to demonstrate that the considered universality
properties allow to perform simple accurate estimates of the EAS
characteristics and that the functional forms are really universal
in terms of energy, primary particle nature and zenith
angle. Certainly, their application to the real experimental
conditions may require some corrections of the coefficients or even of
the functional form of the parametrizations (especially if core
distances different from 1000 meters are considered).

Hence, the \sigrat\ (\denrat) vs \xmaxv\ universality allows to obtain
accurate estimates of the muon signal (muon density) using simple
parametrizations, which are almost independent on the primary particle
nature, primary energy and zenith angle for various types of ground
detectors. Taking into account that the shower universality property
was established for different interaction
models~\cite{giller_univers2005,nerling_univers2005,gora_univers2006,ave_icrc30_univers,lipari_univers2008,lafebre_univers2009},
the proposed approach to muon content derivation should not be
specific only to QGSJET~II and our test calculations with EPOS~1.99
confirm it~(see Fig.~\ref{epos_xmaxv}). In case of EPOS~1.99 the fit
in the form~(\ref{eq:fit}) still provides good description of the
simulated data, but, as expected, fit coefficient are different from
those for QGSJET~II.

\section{Showers at the same zenith angles}
Another universality property follows from the study on showers
arriving at the same zenith angles. In this case the average iron shower
has to cross larger slant distance from \xmax\ to the ground with
respect to the average proton shower and this almost equalizes EM
signals (densities) for both primaries at the observation level in the
wide range of zenith angles (Fig.~\ref{Scos}). For the signal at 1000
meters in the Cherenkov water tanks, notable discrepancies between p
and Fe showers EM components are observed for nearly vertical showers
($\theta<18^{\circ},\ \cos^2(\theta)>0.9$) and very inclined ones
($\theta>63^{\circ},\ \cos^2(\theta)<0.2$). In the first case the
situation is similar to the one considered in Fig.~\ref{DGSlant}: the
path from \xmax\ to the ground for p and Fe showers is almost the
same. For inclined showers the difference is caused by the EM halo
from muon decays and larger number of muons in iron showers brings to
a larger EM halo signal. We normalize \sem\ by primary energy to
cancel the almost linear growth of the signal with the energy in order
to obtain a universal description of \sem\ for all energies.

Looking at the showers at different zenith angles one samples
longitudinal showers profiles, for this reason it is natural to try to
describe the dependence of the EM signal on $\cos(\theta)$ with
Gaisser-Hillas type function, using $\cos(\theta)$ as variable instead
of \xmax:
\begin{multline}
\label{eq:semcos}
\sem(\cos(\theta))/E\ \mathrm{[VEM/EeV]}=\sem^0\left(\frac{\cos(\theta)-c_0}{c_1-c_0}\right)^\alpha\times\\ 
\times\exp\left(\frac{c_1-\cos(\theta)}{\lambda}\right),
\end{multline}
where $\alpha=(c_1-c_0)/\lambda$; $\sem^0$ (signal at maximum), $c_0$
and $c_1$ (cosine of angle at which \sem=$\sem^0$) are fit
parameters. The fit parameters $\sem^0$ and $c_1$ change by less than
10\% and 3\% correspondingly across entire range of energies (when one
makes fits in 15 energy bins $\Delta\log10(E)$[eV]=0.1 from $\logen18.5$
to $\logen20.0$), while $c_0$ changes quite chaotically from $0$ to $-20$
(this causes to change also $\lambda$). We have found that, fixing
$c_0$ (similarly to~\cite{ave_munum_2007}) to any negative value
within this range, we obtain a good universal fit and
$\lambda$ changes in this case by less than 15\%. Finally, we used the
following average values (except $c_0$ that was fixed to $-3$) of the
coefficients $\sem^0=2.53$, $c_0=-3$, $c_1=0.96$, $\lambda=0.012$. The
results of the fit and the difference between the MC simulated EM
signal $\sem^\mathrm{MC}$ and the EM signal obtained from the fit
$\sem^\mathrm{fit}$ are shown in Fig.~\ref{Sdiffcos}. The accuracy of
the reconstruction for all energy bins is shown in Fig.~\ref{SemcosE}
and it is seen that one gets an unbiased estimate of \sem\ with RMS
below 15\% for proton and 13\% for iron showers. Let us note that RMS
of the distributions of relative difference between fit and MC EM
signals reflects shower-to-shower fluctuations and that the
application of fits with coefficients calculated accurately for each
energy bin does not produce any reduction of the RMS.

The same fit procedure was also applied to the density of electrons
$D_e$. In this case good agreement between
p and Fe EAS electron densities was found in the wider angular range
$0-63^\circ$ and $c_0$ was fixed to zero, that gave the
following set of fit parameters: $D_e^0=0.43$ (instead of $\sem^0$),
$c_0=0$, $c_1=0.96$, $\lambda=0.069$. The results of the fit and
electron density reconstruction accuracy are presented in
Figs.~\ref{Ddiffcos},\,\ref{DcosE}. It is seen, that shower-to-shower
fluctuations here are larger compared to the EM signal in water
Cherenkov tanks and while the density estimates are still unbiased,
their spread reaches 25\% and 22\% on average for proton and iron
showers correspondingly.

Finally, our test calculations presented in
Fig.~\ref{epos_cos} demonstrate that the universality of EM signal
(density) dependence on zenith angle holds true also in case of
EPOS~1.99.

\section*{Conclusions}
We have proposed two new EAS universality properties providing two
independent ways to access EM and muon shower contents. We have shown
that these properties can be described with simple parametrizations
which are valid in the wide energy ($\logen18.5-20.0$) and zenith
angle ranges, and independent on the primary particle
nature. Certainly, application of these parametrizations to the real
experimental conditions is not straightforward, but any their
modifications can be easily accomplished and we believe that these
universality properties can be used in hybrid experiments for mass
composition studies, primary and missing energy estimates and for
hadronic interaction model tests. In particular, we would like to
dwell here on the problem of muon excess in the real data compared to
predictions of most of the interaction
models~\cite{abuzayyad_mass2000,engel_icrc2007,ave_munum_2007}. Since
the muon content of EAS is highly model-dependent (see
e.g.~\cite{yakutsk_mu2008,epos_muprod2008}) and the UHECR mass
composition is still unknown, this muon excess can be expressed only
in terms of a relative excess with respect to the prediction of a
given hadronic interaction model for a given primary. In our point of
view primary mass-insensitivity of the presented universality
properties can be used to reveal inconsistencies in the hadronic
interaction models without bias coming from the unknown mass of the
UHECR. One of the possible strategies lies in simultaneous application
of both considered universality properties to the data in order to
reveal hadronic model inconsistencies and, after taking them into
account, to get concording estimates of muon and EM contents. Once it
will be possible to get \sigrat\ vs \xmaxv\ dependence for the real
data (and if the universality~(\ref{eq:fit}) will be observed also
there) it will be clear how one should rescale number of muons
predicted by QGSJET~II or by another model and this correction will be
independent on unknown primary mass. Another possible application of
the property~(\ref{eq:fit}) for muon content derivation in almost
interaction model independent way with its consequent use for primary
mass composition analysis will be presented
elsewhere~\cite{ya_astrop_2009}. And finally, for hybrid detectors
equipped with muon counters probably it will be possible to use
\sigrat\ vs \xmaxv\ universality for determination of depth of shower maximum
taking advantage of 100\% ground array duty cycle in respect to 10\%
one of the fluorescence telescopes, but certainly this problem
requires dedicated study.

\subsection*{Acknowledgements}
We are very grateful to Maximo Ave and Fabian Schmidt for kind
permission to use their GEANT~4 lookup tables in our calculations of
signal from different particles in Auger water tanks.

\end{document}